\documentclass[11pt]{article}

\usepackage[final]{acl}

\usepackage{times}
\usepackage{latexsym}
\usepackage{amsmath}
\usepackage{amssymb}
\usepackage{enumitem}

\usepackage[T1]{fontenc}

\usepackage[utf8]{inputenc}
\usepackage{booktabs}   
\usepackage{multirow}   
\usepackage{caption}    
\usepackage{hyperref}

\usepackage{microtype}

\usepackage{inconsolata}

\usepackage{graphicx}

\usepackage{tcolorbox}

\usepackage{listings}
\usepackage{xcolor}
\usepackage{adjustbox}

\title{CypherMAS: Benchmarking Routing Reliability in LLM-Based Multi-Agent Systems}

\author{Danil Kostromin\textsuperscript{\rm 1, 2}, Andrey Sakhovskiy\textsuperscript{\rm 1},
Semen Budennyy\textsuperscript{\rm 1,3} 
\\
\textsuperscript{\rm 1} Sber AI
\textsuperscript{\rm 2} RTU MIREA
\textsuperscript{\rm 3} AIRI \\
    {\small
    \textbf{Correspondence:} \href{mailto:andrey.sakhovskiy@gmail.com}{andrey.sakhovskiy@gmail.com}}}

\begin{document}
\maketitle
\begin{abstract}
Multi-agent systems (MAS) built on large language models (LLMs) distribute reasoning across specialized agents that must exchange messages and invoke tools reliably. Existing evaluations typically couple communication with open domain tasks, making it difficult to determine whether a failure arises from reasoning, malformed tool use, lost messages, or forgotten requests.
Prior benchmarks do not isolate communication execution or measure routing redundancy against a known optimal path.
We introduce CypherMAS, a synthetic benchmark that isolates communication and routing in fixed multi-agent graphs. Agents in chain, star, bipartite, or ring graphs route cyphered messages to remote peers holding the matching decoders; since the optimal route is known, redundant LLM calls can be measured and routing failures may be easily tracked. Across eleven open-weight models from 4B to 753B, small models fail at routing and tool use, dense models above 27B degrade with diameter, and successful runs exceed minimum LLM calls. These findings identify communication reliability as a distinct scaling bottleneck for multi-agent systems and provide a controlled diagnostic for complex multi-agent systems. The source code is freely available\footnote{\url{https://github.com/Andoree/CypherMAS}}.

\end{abstract}

\section{Introduction}

\begin{figure*}[t]
\centering
\includegraphics[width=0.9\textwidth]{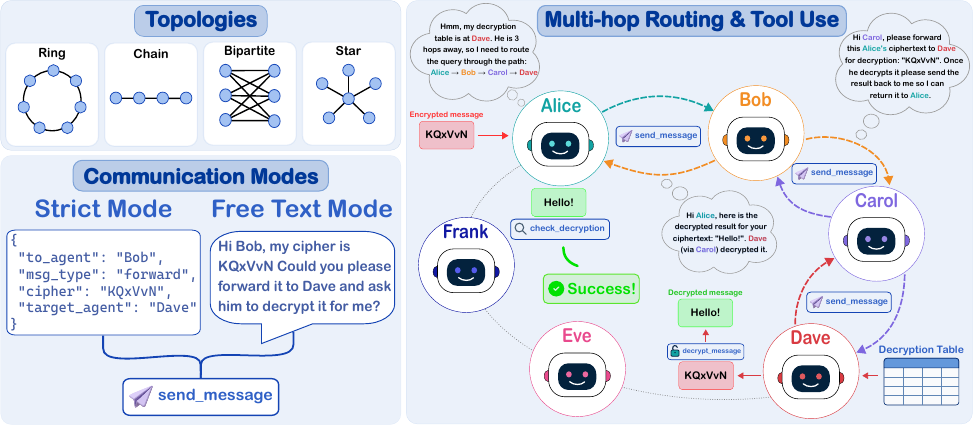} 
\caption{Overview of CypherMAS. \textbf{Left:} agents operate on one of four fixed topologies and communicate through typed messages (\textit{Strict Mode}) or free-form natural language (\textit{Free Text Mode}). \textbf{Right:} Alice routes her ciphertext along the shortest path to Dave, who holds the matching decoder three hops away (dashed arrows); the plaintext returns along the reverse path (solid arrows) and is verified with \texttt{check\_decryption}. All agents pursue the same objective concurrently while forwarding and decrypting for peers. As the graph and decoder placement are known, the shortest request--response path defines the minimum number of LLM calls, and any additional call measures routing redundancy.}
\label{fig:teaser}
\vspace{-0.5cm}
\end{figure*}

LLM-based multi-agent systems (MAS) distribute reasoning across specialized agents that exchange messages, invoke tools, and track pending requests~\citep{MultiAGentDebateICML-2024,MDAgents-NeurIPS2024,zhou-etal-2025-reso,KG4Diagnosis-AAAI2025,qiang2025mlesmithscalingmletasks,su-etal-2025-many,tan-etal-2026-scitrue}.
Scaling to dozens of agents introduces failure points absent from single-agent inference: e.g., tool call hallucinations and pending requests loss.
Recent work constructs task-specific MAS pipelines or automates their topologies~\citep{li2023camel,Autogen,hong2024metagpt,MDAgents-NeurIPS2024,GPTSwarmICML2024,GDesignerICML2025,zhou-etal-2025-reso}.
Recent failure attribution research reduces MAS failure to a binary outcome~\citep{MAST-Why-MAS-fail-NIPS2025,AgentTracer-ICLR2026-but-arxiv,WhoWhenMAS-ICML2025,chen-etal-2026-seeing}: a MAS run either produces the correct final answer or fails.
This formulation neglects communication overhead. A system may succeed while routing messages through redundant loops, masking malformed tool calls with exception handlers, or issuing several times the minimum number of transmissions, thus increasing inference cost.
Quantifying this overhead requires a known minimum communication bound per agent. However, the existing benchmarks evaluate on open-ended tasks, providing no theoretical redundancy estimates~\citep{MultiAGentDebateICML-2024,Autogen,hong2024metagpt,MDAgents-NeurIPS2024,KG4Diagnosis-AAAI2025,GPTSwarmICML2024,qiang2025mlesmithscalingmletasks,zhou-etal-2025-reso,su-etal-2025-many,tan-etal-2026-scitrue}. A  communication reduction may merely reflect that a task lacks genuine multi-agent necessity, rather than demonstrating improved routing efficiency.

We introduce CypherMAS, a synthetic benchmark that provides exact lower-bound routing diagnostics at the level of individual agents. The benchmark replaces a shared group goal with individual objectives that require cooperation. As seen in Figure~\ref{fig:teaser}, each agent receives an encrypted message and a decryption table for another agent's message. The table required to decrypt its own message is held by a peer. An agent succeeds only if its encrypted message reaches the correct peer, that peer decrypts it, and the decrypted text returns to the owner and is verified. All agents pursue this objective concurrently over a fixed, fully observable graph. Agents communicate only with direct neighbours, and message delivery, decryption, and verification occur through tool calls. As the graph and the location of each required decryption table are known, the shortest request and response path is known. The benchmark therefore provides a ground-truth lower bound on the number of LLM calls required by each configuration. Additional calls measure routing redundancy. Failed calls can be attributed to execution rather than task reasoning.
\emph{CypherMAS} evaluates two communication regimes.
i. \emph{Strict Mode} uses structured tool calls with fixed fields, isolating routing and tool execution from free-form natural language requests.
ii. \emph{Free Text Mode} uses unrestricted natural language messages, testing request formulation, forwarding, and response parsing.
Through \emph{CypherMAS} design, we explore three  research questions:

\begin{itemize}[itemsep=-1.25pt, topsep=-1.5pt]
    \item \textbf{RQ1.} Can current LLMs route messages and invoke tools reliably in fixed graphs depending on model scale and architecture?
    \item \textbf{RQ2.} How much communication overhead do models incur beyond the theoretical minimum as graph size and diameter increase?
    \item \textbf{RQ3.} Does strict tool-based communication produce fewer routing and tool errors?
\end{itemize}

Our contributions are:
\begin{itemize}[itemsep=-1.75pt,topsep=-1.5pt]

\item \textsc{CypherMAS}, a multi-agent benchmark that provides exact routing lower bounds, allowing deterministic classification of every message as on-path, misrouted, dropped, or duplicated.
\item A taxonomy of communication failures, including dropped requests, loops, and malformed calls, analyzed by graph size, diameter, and model architecture.
\item An empirical study of eleven open-weight models ranging from 4B to 753B parameters that identifies where routing and tool execution break down as graphs grow.
\end{itemize}

\begin{table*}[t!]
  \centering
  \setlength{\tabcolsep}{3pt} 
  

  \resizebox{\textwidth}{!}{%
  \begin{tabular}{@{} l *{4}{cc} @{}}
    \toprule
    \multirow{2}{*}{\textbf{LLM}} 
    & \multicolumn{2}{c}{\textbf{Chain}} 
    & \multicolumn{2}{c}{\textbf{Star}} 
    & \multicolumn{2}{c}{\textbf{Bipartite Graph}} 
    & \multicolumn{2}{c}{\textbf{Ring}} \\
    
    \cmidrule(lr){2-3} \cmidrule(lr){4-5} \cmidrule(lr){6-7} \cmidrule(lr){8-9}
    
    & \textit{SR} & \textit{LLM Calls} 
    & \textit{SR} & \textit{LLM Calls} 
    & \textit{SR} & \textit{LLM Calls} 
    & \textit{SR} & \textit{LLM Calls} \\
    \midrule
    \multicolumn{9}{c}{\textbf{Strict (Tool-Based) Communication}} \\ \hline 
    
    Qwen3.5-4B      & 0.03$_{\pm 0.02}$ & 1692$_{\pm 619}$ & 0.12$_{\pm 0.00}$ & 174$_{\pm 70}$ & 0.05$_{\pm 0.06}$ & 172$_{\pm 36}$ & 0.00$_{\pm 0.00}$ & 2049$_{\pm 1}$ \\
    Qwen3.5-9B      & 0.01$_{\pm 0.02}$ & 874$_{\pm 258}$ & 0.01$_{\pm 0.02}$ & 39$_{\pm 24}$ & 0.04$_{\pm 0.04}$ & 95$_{\pm 42}$ & 0.00$_{\pm 0.00}$ & 471$_{\pm 124}$ \\
    Qwen3.6-27B      & 0.36$_{\pm 0.04}$ & 549$_{\pm 38}$ & 1.00$_{\pm 0.00}$ & 96$_{\pm 0}$ & 0.99$_{\pm 0.02}$ & 99$_{\pm 1}$ & 0.57$_{\pm 0.10}$ & 624$_{\pm 66}$ \\
    Qwen3.6-35B-A3B      & 0.07$_{\pm 0.02}$ & 362$_{\pm 29}$ & 0.27$_{\pm 0.29}$ & 78$_{\pm 45}$ & 0.64$_{\pm 0.32}$ & 110$_{\pm 11}$ & 0.00$_{\pm 0.00}$ & 487$_{\pm 140}$ \\
    Qwen3.8-27B      & 0.55$_{\pm 0.08}$ & 553$_{\pm 51}$ & 1.00$_{\pm 0.00}$ & 96$_{\pm 0}$ & 1.00$_{\pm 0.00}$ & 120$_{\pm 34}$ & 0.75$_{\pm 0.12}$ & 702$_{\pm 86}$ \\
    Gemma4-31B      & 0.85$_{\pm 0.02}$ & 601$_{\pm 4}$ & 1.00$_{\pm 0.00}$ & 96$_{\pm 0}$ & 1.00$_{\pm 0.00}$ & 100$_{\pm 0}$ & 0.89$_{\pm 0.08}$ & 597$_{\pm 28}$ \\
    Devstral-2-24B      & 0.01$_{\pm 0.02}$ & 177$_{\pm 31}$ & 0.39$_{\pm 0.31}$ & 105$_{\pm 30}$ & 0.27$_{\pm 0.18}$ & 214$_{\pm 85}$ & 0.00$_{\pm 0.00}$ & 218$_{\pm 37}$ \\
    gpt-oss-20b      & 0.01$_{\pm 0.02}$ & 105$_{\pm 21}$ & 0.04$_{\pm 0.00}$ & 26$_{\pm 1}$ & 0.01$_{\pm 0.02}$ & 28$_{\pm 1}$ & 0.00$_{\pm 0.00}$ & 110$_{\pm 18}$ \\
    gpt-oss-120b      & 0.05$_{\pm 0.02}$ & 159$_{\pm 5}$ & 0.08$_{\pm 0.04}$ & 30$_{\pm 4}$ & 0.09$_{\pm 0.05}$ & 33$_{\pm 3}$ & 0.00$_{\pm 0.00}$ & 142$_{\pm 4}$ \\
    DeepSeek-V4-Flash      & 0.87$_{\pm 0.06}$ & 581$_{\pm 8}$ & 1.00$_{\pm 0.00}$ & 96$_{\pm 0}$ & 1.00$_{\pm 0.00}$ & 100$_{\pm 0}$ & 0.69$_{\pm 0.21}$ & 600$_{\pm 60}$ \\
    GLM-5.2      & 0.91$_{\pm 0.08}$ & 1248$_{\pm 719}$ & 1.00$_{\pm 0.00}$ & 96$_{\pm 0}$ & 1.00$_{\pm 0.00}$ & 100$_{\pm 0}$ & 0.97$_{\pm 0.02}$ & 729$_{\pm 66}$ \\

    \midrule
    \multicolumn{9}{c}{\textbf{Free Form Communication}} \\ \hline 
    Qwen3.5-4B      & 0.04$_{\pm 0.00}$ & 1739$_{\pm 271}$ & 0.40$_{\pm 0.42}$ & 761$_{\pm 506}$ & 0.36$_{\pm 0.08}$ & 637$_{\pm 669}$ & 0.00$_{\pm 0.00}$ & 2061$_{\pm 1}$ \\
    Qwen3.5-9B      & 0.05$_{\pm 0.05}$ & 1480$_{\pm 1001}$ & 0.48$_{\pm 0.42}$ & 175$_{\pm 188}$ & 0.33$_{\pm 0.26}$ & 124$_{\pm 36}$ & 0.00$_{\pm 0.00}$ & 2074$_{\pm 11}$ \\
    Qwen3.6-27B      & 0.37$_{\pm 0.12}$ & 1156$_{\pm 829}$ & 0.85$_{\pm 0.14}$ & 88$_{\pm 7}$ & 0.97$_{\pm 0.02}$ & 99$_{\pm 1}$ & 0.47$_{\pm 0.22}$ & 1073$_{\pm 858}$ \\
    Qwen3.6-35B-A3B      & 0.13$_{\pm 0.09}$ & 628$_{\pm 378}$ & 0.39$_{\pm 0.53}$ & 50$_{\pm 40}$ & 0.96$_{\pm 0.00}$ & 135$_{\pm 43}$ & 0.01$_{\pm 0.02}$ & 1885$_{\pm 300}$ \\
    Qwen3.8-27B      & 0.96$_{\pm 0.04}$ & 646$_{\pm 5}$ & 1.00$_{\pm 0.00}$ & 98$_{\pm 3}$ & 1.00$_{\pm 0.00}$ & 112$_{\pm 14}$ & 0.93$_{\pm 0.06}$ & 720$_{\pm 58}$ \\
    Gemma4-31B      & 0.87$_{\pm 0.14}$ & 835$_{\pm 408}$ & 0.99$_{\pm 0.02}$ & 95$_{\pm 1}$ & 0.83$_{\pm 0.30}$ & 97$_{\pm 5}$ & 0.77$_{\pm 0.15}$ & 1078$_{\pm 191}$ \\
    Devstral-2-24B      & 0.01$_{\pm 0.02}$ & 710$_{\pm 1162}$ & 0.53$_{\pm 0.39}$ & 66$_{\pm 38}$ & 0.45$_{\pm 0.42}$ & 100$_{\pm 49}$ & 0.00$_{\pm 0.00}$ & 713$_{\pm 1158}$ \\
    gpt-oss-20b      & 0.00$_{\pm 0.00}$ & 81$_{\pm 3}$ & 0.00$_{\pm 0.00}$ & 26$_{\pm 1}$ & 0.00$_{\pm 0.00}$ & 25$_{\pm 2}$ & 0.00$_{\pm 0.00}$ & 77$_{\pm 5}$ \\
    gpt-oss-120b      & 0.00$_{\pm 0.00}$ & 174$_{\pm 23}$ & 0.04$_{\pm 0.00}$ & 28$_{\pm 2}$ & 0.03$_{\pm 0.02}$ & 33$_{\pm 6}$ & 0.00$_{\pm 0.00}$ & 169$_{\pm 34}$ \\
    DeepSeek-V4-Flash      & 0.95$_{\pm 0.02}$ & 1139$_{\pm 789}$ & 1.00$_{\pm 0.00}$ & 96$_{\pm 0}$ & 1.00$_{\pm 0.00}$ & 104$_{\pm 5}$ & 0.99$_{\pm 0.02}$ & 691$_{\pm 28}$ \\
    GLM-5.2      & 0.96$_{\pm 0.00}$ & 573$_{\pm 9}$ & 1.00$_{\pm 0.00}$ & 96$_{\pm 1}$ & 1.00$_{\pm 0.00}$ & 246$_{\pm 5}$ & 1.00$_{\pm 0.00}$ & 947$_{\pm 21}$ \\

    \bottomrule
  \end{tabular}}\caption{Performance evaluation of supported LLMs across various graph topologies with 25 agents. \emph{SR} stands for success rate. Each metric value is averaged across 3 independent MAS runs and $\pm$ denotes standard deviation. The theoretical minimum of LLM calls is: Chain - 624, Star - 96, Bipartite - 100, Ring - 600.}\label{tab:llm-topologies}
  \vspace{-0.5cm}

\end{table*}


\begin{figure*}[h]\centering

\includegraphics[width=\textwidth]{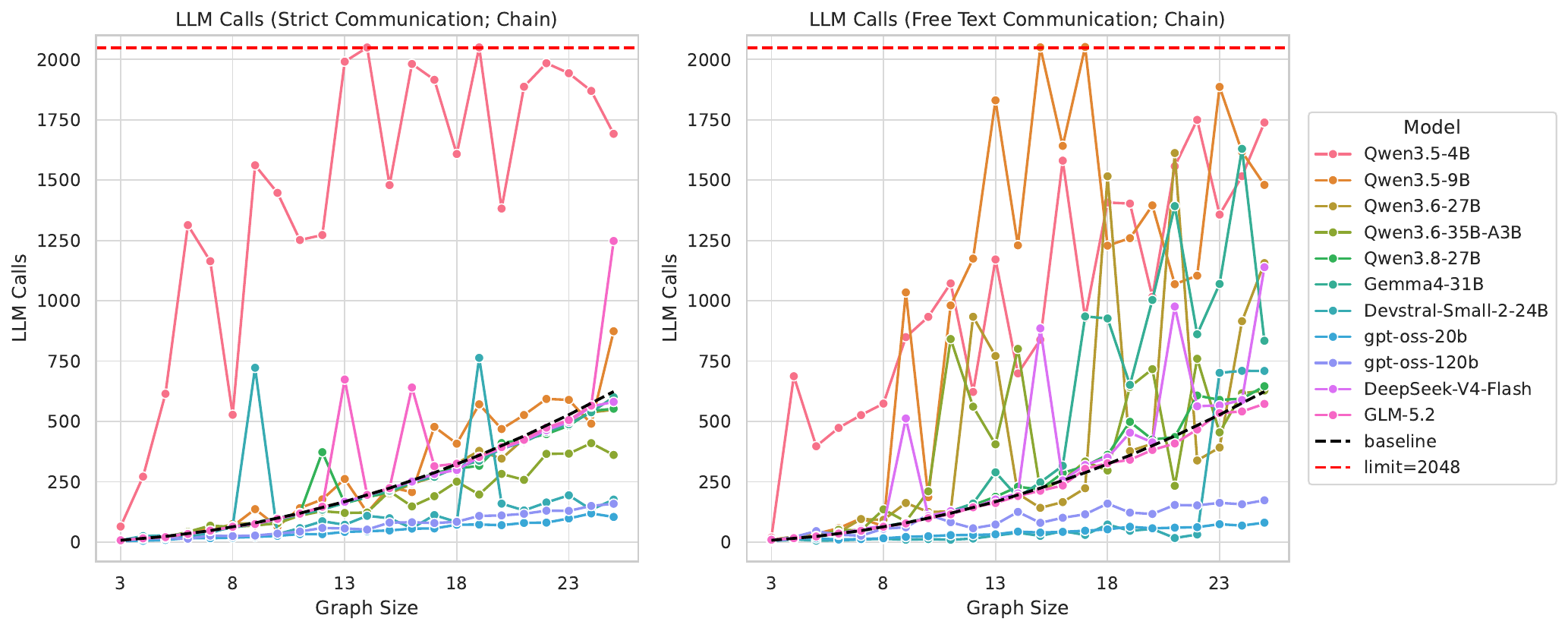}

\vspace{0.3cm}

\includegraphics[width=\textwidth]{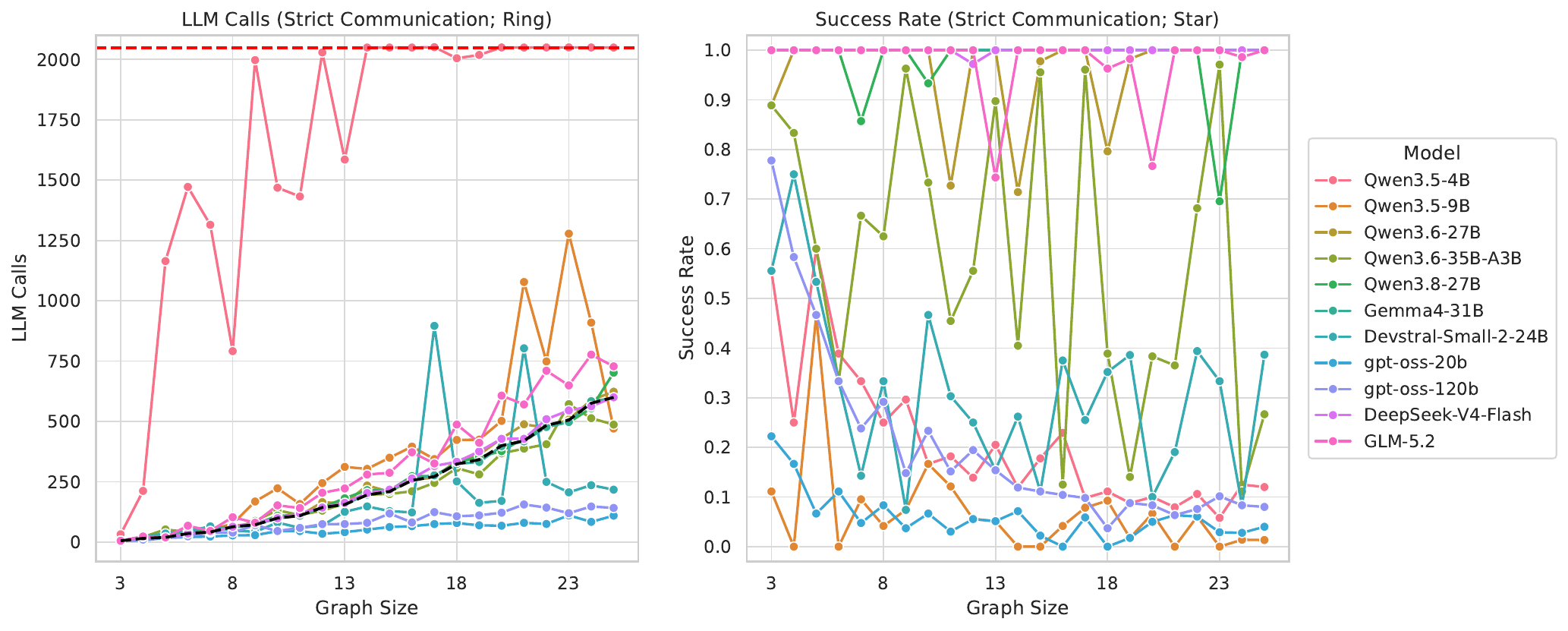}

\caption{Representative scaling behaviors as graph size $N$ grows from 3 to 25. From left to right: LLM calls on Chain in Strict mode, where \textit{GLM-5.2} issues roughly twice the theoretical minimum due to duplicated messages; LLM calls on Chain in Free Text mode, where the same model stays near the baseline; LLM calls on Ring in Strict mode, where small models hit the 2{,}048-call cap; success rate on Star in Strict mode, where bounded diameter removes the scaling effect for capable models. Dashed horizontal lines mark the theoretical minimum call count (\emph{baseline}). Full plots for all topologies and metrics are provided in Appendix~\ref{appx:appendix-results}.}
\label{fig:scaling-subset}
\vspace{-0.5cm}
\end{figure*}


\section{Related Work}

\paragraph{LLM-based Multi-Agent Systems.}
Existing research on MAS construction automation has focused primarily on three key directions. Handcrafted systems fix roles and communication protocols a priori~\cite{li2023camel,Autogen,hong2024metagpt,MDAgents-NeurIPS2024,KG4Diagnosis-AAAI2025,tan-etal-2026-scitrue,su-etal-2025-many,SAPO-10.1145/3786335.3813128-2026}, as seen in MetaGPT~\citep{hong2024metagpt}, AutoGen~\citep{Autogen}, and CAMEL~\citep{li2023camel}. Partially automated systems optimize individual components, such as agent selection~\cite{Dylan2024,zhou-etal-2025-reso,SocialDropout-SIGIR2026}, prompt design~\citep{PromptBreeder-ICLR-2024,EvoPromptingICLR-2024,chu-etal-2025-exploring}, or inter-agent topology~\citep{GPTSwarmICML2024,GDesignerICML2025,SafeSieveAAAI2026}.

While prior research explores communication redundancy and topology optimisation (e.g., AgentPrune~\cite{AgentPruneICLR2025} or G-Designer~\cite{GDesignerICML2025}), these methods typically evaluate procedural tasks where the theoretical minimum communication is unknown. Consequently, they measure redundancy indirectly via changes in a shared final outcome. CypherMAS complements this by providing a ground-truth lower bound, enabling exact per-hop analysis.

 \begin{table*}[t!]
\centering
\setlength{\tabcolsep}{8pt} 
\resizebox{0.9\textwidth}{!}{ 
\begin{tabular}{llcccc}

\toprule
\textbf{Topology} & \textbf{Model} & \textbf{\shortstack{Misrouted Msgs}} & \textbf{\shortstack{Dropped Request}} & \textbf{\shortstack{Duplicate Msg}} & \textbf{\shortstack{Total Msgs}} \\
\midrule

\multirow{11}{*}{Chain} 
 & Qwen3.5-4B      & 678 (40.05\%) & 13 (0.75\%) & 1406 (83.08\%) & 1692 \\
 & Qwen3.5-9B      & 286 (32.72\%) & 5 (0.57\%) & 554 (63.42\%) & 874 \\
 & Qwen3.6-27B      & 23 (4.13\%) & 1 (0.24\%) & 23 (4.25\%) & 549 \\
 & Qwen3.6-35B-A3B      & 58 (15.94\%) & 9 (2.58\%) & 58 (16.13\%) & 362 \\
 & Qwen3.8-27B      & 19 (3.44\%) & 0 (0.00\%) & 0 (0.00\%) & 553 \\
 & Gemma4-31B      & 3 (0.44\%) & 0 (0.00\%) & 1 (0.17\%) & 601 \\
 & Devstral-2-24B      & 2 (1.32\%) & 3 (1.70\%) & 21 (11.70\%) & 177 \\
 & gpt-oss-20b      & 2 (1.91\%) & 24 (22.93\%) & 18 (17.20\%) & 105 \\
 & gpt-oss-120b      & 1 (0.84\%) & 22 (14.02\%) & 20 (12.34\%) & 159 \\
 & DeepSeek-V4-Flash      & 0 (0.00\%) & 1 (0.17\%) & 4 (0.69\%) & 581 \\
 & GLM-5.2      & 51 (4.06\%) & 0 (0.00\%) & 562 (45.03\%) & 1248 \\
\midrule



\multirow{11}{*}{Ring} 
 & Qwen3.5-4B      & 349 (17.01\%) & 61 (2.96\%) & 1684 (82.16\%) & 2049 \\
 & Qwen3.5-9B      & 119 (25.34\%) & 8 (1.63\%) & 194 (41.26\%) & 471 \\
 & Qwen3.6-27B      & 23 (3.74\%) & 1 (0.21\%) & 35 (5.56\%) & 624 \\
 & Qwen3.6-35B-A3B      & 115 (23.53\%) & 12 (2.39\%) & 111 (22.85\%) & 487 \\
 & Qwen3.8-27B      & 42 (5.98\%) & 0 (0.00\%) & 85 (12.11\%) & 702 \\
 & Gemma4-31B      & 3 (0.50\%) & 0 (0.00\%) & 8 (1.34\%) & 597 \\
 & Devstral-2-24B      & 0 (0.00\%) & 3 (1.53\%) & 39 (18.07\%) & 218 \\
 & gpt-oss-20b      & 7 (6.06\%) & 22 (20.00\%) & 16 (14.24\%) & 110 \\
 & gpt-oss-120b      & 3 (2.11\%) & 27 (19.25\%) & 30 (20.89\%) & 142 \\
 & DeepSeek-V4-Flash      & 6 (1.03\%) & 1 (0.17\%) & 24 (4.00\%) & 600 \\
 & GLM-5.2      & 4 (0.59\%) & 0 (0.05\%) & 3 (0.37\%) & 729 \\
\bottomrule

\end{tabular}
}
\caption{Message-level communication errors for all eleven models on the Chain and Ring topologies; the corresponding Star and Bipartite results are reported in Appendix~\ref{appx:appendix-errors} (Table~\ref{tab:msg-errors-sb}). \textit{Misrouted Msgs} combines ciphertexts sent away from the decoder holder and decryptions sent away from the ciphertext owner. \textit{Dropped Request} indicates unactioned forward or decrypt requests. \textit{Duplicate Msg} indicates looped or repeated messages. \textit{Total Msgs} reports the average total messages sent per run.}
\label{tab:message-level-errors}
\vspace{-0.5cm}
\end{table*}

\paragraph{Benchmarks for LLM Agents.}
Current MAS are frequently evaluated on general-purpose tasks, e.g., mathematics or code generation~\cite{GDesignerICML2025,GPTSwarmICML2024,AssembleCrewAAAI2026-ARG-Designer}.  Dedicated agent benchmarks address tool use and reasoning in complex environments \citep{AgentBench2024-ICLR, WebArena2024-ICLR, SWE-Agent-arxiv-2024, GTA-NeurIPS2024, li-etal-2023-api}.  They primarily focus on single agents operating within complex
execution pipelines.
These benchmarks couple tool execution
with semantic reasoning, which prevents the isolation of
mechanical execution errors from open-domain reasoning capabilities.
Multi-agent benchmarks extend to collaborative settings~\citep{zhu-etal-2025-multiagentbench,lu-etal-2026-octotools}. 
All these datasets lack agentic specificity and typically provide no theoretical guarantees on whether the multi-agent approach must be preferred over a single LLM and enable no algorithmic analysis of routing overhead.



\paragraph{Failure attribution and MAS resilience.}
One line of MAS reliability attributes failures to specific agents and steps.
MAST~\cite{MAST-Why-MAS-fail-NIPS2025},
Who\&When~\cite{WhoWhenMAS-ICML2025}, and
AgentTracer~\cite{AgentTracer-ICLR2026-but-arxiv}
localise failures to specific agents and steps.
Alternatively,~\citet{DBLP:conf/icml/0001ZJZC0YLS25} and~\citet{miao-etal-2026-blindguard} evaluate topology degradation under faulty agents. ResMAS~\citep{DBLP:conf/aaai/ZhouLLSWSJX26} optimizes topologies and prompts to increase resilience.
However, existing research reduces system quality to a binary task outcome: all agents solve a single shared task collaboratively. 
Redundant routing loops, silently handled tool errors, and unnecessary LLM invocations do not alter the final answer and therefore go undetected.

%


\section{CypherMAS Benchmark}

CypherMAS replaces open-ended collaboration with a deterministic routing task. The task is analogous to a set of distributed locked boxes: each agent holds a locked box and a key for another agent's box. No agent can open its own box. To obtain the \emph{plaintext}, an agent must route its \emph{ciphertext} through the communication graph to the peer that holds the matching \emph{decoder}, and the plaintext must then return along the reverse path. Because the graph is fully observable and each agent knows the peer that holds its decoder, the shortest agent--decoder path is computable in advance. Therefore, errors can be studied as failures in message handling, tool invocation, and protocol execution rather than as failures in planning under semantic uncertainty.

\subsection{Task Formulation}

\paragraph{Environment.}
The environment is a fixed undirected graph $\mathcal{G} = (\mathcal{V}, \mathcal{E})$, where $\mathcal{V} = \{A_1, A_2, \dots, A_N\}$ is the set of agents; $\mathcal{E}$ is the set of edges that define the agentic topology: if two agents are connected to an edge, they may send each other queries directly. Agent $A_{i}$ can exchange messages only with direct neighbors $\mathcal{N}(A_i)$. Decoder assignment is a fixed derangement $\pi$ over $\{1, \dots, N\}$: agent $A_{\pi(i)}$ holds the decoder for ciphertext $C_{i}$ of agent $A_i$, and $\pi(i) \neq i$. 

The benchmark enforces five rules:
\begin{enumerate}[itemsep=-3.25pt, topsep=1.5pt]
    \item $G$ is fixed and visible to all agents.
    \item Each agent receives a ciphertext and a decoder. No agent can decrypt its own ciphertext.
    \item Each agent knows $A_{\pi(i)}$, the holder of the decoder for its own ciphertext.
    \item Agents communicate only with direct neighbors. Global broadcasting is not allowed.
    \item Decryption and verification are performed only through the provided tools.
\end{enumerate}

\paragraph{Initialization.}
Each agent $A_i$ receives: (i) a unique identifier and the graph $\mathcal{G}$; (ii) its ciphertext $C_i$; (iii) a decoder for $C_{\pi(i)}$; (iv) routing metadata identifying $A_{\pi(i)}$, the decoder holder for for $C_i$.

\paragraph{Objective.}
Agent $A_i$ must recover and verify the plaintext $P_i$ of its own ciphertext $C_i$. If $A_{\pi(i)} \in \mathcal{N}(A_i)$, $A_i$ sends $C_i$ directly. Otherwise, $A_i$ routes $C_i$ through intermediate agents along a path to $A_{\pi(i)}$. The plaintext then returns along the reverse path. At the same time, each agent must decrypt incoming ciphertexts for which it holds the decoder and return the resulting plaintext to the requesting owner. The global task is complete only when all agents have verified their plaintexts.

\paragraph{Design rationale.}
The task separates communication execution from semantic reasoning through three modeling choices. First, \emph{knowledge incompleteness} makes communication necessary: $A_i$ cannot succeed without $A_{\pi(i)}$, so no agent can solve its task in isolation. Second, \emph{tool-mediated state changes} make execution errors observable: forwarding, decryption, and verification occur through typed calls, which separates malformed tool use from reasoning over valid inputs. Third, \emph{fixed topology} makes redundancy measurable. Let $d_i = d(A_i, A_{\pi(i)})$
be the shortest-path distance between $A_i$ and its decoder holder in $\mathcal{G}$. A successful request--response cycle for $C_i$ requires at least $2d_i$ neighbor-to-neighbor transmissions: $d_i$ forward transmissions and $d_i$ return transmissions. Because each transmission corresponds to a \texttt{send\_message} invocation, $2d_i$ is a lower bound on message-level LLM calls for that agent. Summing over agents gives a run-level lower bound, and deviations from this bound quantify routing redundancy.

\subsection{Agent Interface and Execution}

\paragraph{Tools.}
Let $A_i$ denote the calling agent. Agents interact through three tools:
\begin{itemize}[itemsep=-2.5pt, topsep=1.25pt]
    \item \texttt{send\_message}$(A_j, m)$ sends message $m$ to agent $A_j$. The call is valid only if $A_j \in \mathcal{N}(A_i)$. If $A_j \notin \mathcal{N}(A_i)$, the tool returns an error. Invoking this tool ends the agent's current execution hop, preventing further reasoning or tool calls in the same turn. Agents may issue multiple parallel \texttt{send\_message} calls within one step to send distinct ciphertexts.

    \item \texttt{decrypt\_message}$(C_k)$ applies the decoder assigned to $A_i$.
      The call returns the correct plaintext iff $k = \pi(i)$, otherwise returning an incorrect plaintext.
      Since $\pi(i) \neq i$, no agent can decrypt its own ciphertext. Calls with $k \neq \pi(i)$ are recorded as decryption errors.
    
    \item \texttt{check\_decryption}$(\hat{P}_i)$ verifies task completion for $A_i$. If $\hat{P}_i = P_i$, the tool returns acceptance and completes the objective of $A_i$. If $\hat{P}_i \neq P_i$, the tool returns rejection and identifies $A_{\pi(i)}$, preventing infinite guessing loops and providing explicit routing feedback.
\end{itemize}

\paragraph{Communication modes.}
We evaluate two execution modes to isolate distinct aspects of LLM communication.
In \textbf{Strict Mode}, agents bypass natural language generation for routing. They invoke \texttt{send\_message} using strictly typed actions:
\begin{itemize}[itemsep=-2.25pt, topsep=2pt]
    \item \texttt{decrypt\_request}$(A_o, C_o)$: requests decryption of $C_o$ on behalf of $A_o$.
    \item \texttt{decrypt\_response}$(A_o, P_o)$: returns the decrypted plaintext $P_o$ toward the owner $A_o$.
    \item \texttt{forward}$(A_o, C_o, A_{\text{next}})$: routes $C_o$ to the next hop $A_{\text{next}}$ on the shortest path toward $A_{\pi(o)}$.
\end{itemize}
The system formats these structured arguments into a standardized payload before delivery. This mode enforces a fixed message schema and isolates routing errors from natural-language variation.

In \textbf{Free Text Mode}, $m$ is an arbitrary string. The agent selects a target neighbor $A_j \in \mathcal{N}(A_i)$ and sends a free-form natural-language query. This mode tests request formulation, routing instructions, and response parsing without typed structural scaffolding. The complete system prompts for both modes are provided in Appendix~\ref{appx:prompts}. For examples of two communication modes, see Appx.~\ref{appx:appendix-messages}.

\paragraph{Execution protocol.}
Execution is synchronous and round-based. A central message queue stores all inter-agent messages. At the start of round $T$, each agent receives messages sent by neighbors in round $T-1$.
Within a round, an execution hop processes the incoming queue, generates internal reasoning, and issues tool calls. 
The hop ends when the agent makes one or more \texttt{send\_message} calls.

\subsection{Graph Topologies and Scaling}

\paragraph{Topologies.}
\emph{CypherMAS} uses four predefined undirected graphs to isolate distinct structural effects on routing. \textbf{Chain} and \textbf{Ring} explore communication over growing distances: a Chain is a linear sequence where intermediate agents the diameter grows linearly with $N$. Ring closes the loop so that every agent has exactly two neighbors and, under maximum-distance decoder placement, all request paths have equal length $N/2$. \textbf{Star} and \textbf{Bipartite} bound the routing distance at two hops regardless of $N$: a Star routes all traffic through a single hub that must process high concurrent throughput, while a Bipartite graph restricts edges to connections between two disjoint partitions.

\paragraph{Scaling and Decoder Placement.}

To enforce multi-hop communication and ciphertext routing between distant hops in all topologies, we maximize the topological distance an agent $A_i$ and its decoder holder $A_{\pi(i)}$.
Let $d(u, v)$ denote the shortest path distance between agents $u$ and $v$ in $G$. We construct a $N \times N$ cost matrix $D$, where $D_{i,j} = -d(A_i, A_j)$ for $i \neq j$ and $D_{i,i} = \infty$ to prevent self-assignment. The benchmark solves an assignment problem on $D$ using the Hungarian algorithm resulting in the optimal permutation $\pi^*$:
\begin{equation}
    \pi^* = \arg\max_{\pi} \sum_{i=1}^{N} d(A_i, A_{\pi(i)})
\end{equation}

The resulting distance behaviour is topology-specific. Chain and Ring scale the diameter linearly as the $N$ grows. Star and Bipartite bound routing distance at two hops for all $N$, so the distance is constant is 2  regardless of graph size.



\subsection{Error Types}

\textbf{Strict mode} enables algorithmic analysis and deterministic classification of execution failures. \emph{CypherMAS} diagnoses  failure modes classifies into message- and decryption-centric.

\paragraph{Message errors.}
Message errors count as invalid \texttt{send\_message} calls. We record three types of error: i. \textbf{Misdirected cipher}, where a ciphertext is sent via a non-shortest path to the decoder holder; ii. \textbf{Misdirected decryption}, where a plaintext is sent via a non-shortest path to the cypher owner; iii. \textbf{Corrupted cipher}, where the ciphertext argument differs from the exact received ciphertext.

\paragraph{Decryption errors.}
Decryption errors count incorrect \texttt{decrypt\_message} calls. We define two error types: i. \textbf{Wrong decryption}, where an agent applies \texttt{decrypt\_message} to a ciphertext its decoder cannot decrypt; ii. \textbf{Unsent decryption}, where an agent decrypts a ciphertext but does not invoke \texttt{send\_message} to return the plaintext.

\section{Experimental Setup}
\begin{table*}[t!]
\centering
\small
\setlength{\tabcolsep}{1.75pt}
\begin{tabular}{llcccccc}

\toprule
\textbf{Topology} & \textbf{Model} & \shortstack{Completed} & \shortstack{Init Fail} & \shortstack{Transit to Dec} & \shortstack{Transit to Own} & \shortstack{Decoder Stuck} & \shortstack{Verify Stuck} \\
\midrule

\multirow{11}{*}{Chain} 
 & Qwen3.5-4B      & 0.03$_{\pm 0.02}$ & 0.04$_{\pm 0.00}$ & 0.69$_{\pm 0.02}$ & 0.00$_{\pm 0.00}$ & 0.01$_{\pm 0.02}$ & 0.23$_{\pm 0.02}$ \\
 & Qwen3.5-9B      & 0.01$_{\pm 0.02}$ & 0.05$_{\pm 0.06}$ & 0.67$_{\pm 0.06}$ & 0.08$_{\pm 0.04}$ & 0.09$_{\pm 0.02}$ & 0.09$_{\pm 0.06}$ \\
 & Qwen3.6-27B      & 0.36$_{\pm 0.04}$ & 0.00$_{\pm 0.00}$ & 0.09$_{\pm 0.10}$ & 0.55$_{\pm 0.13}$ & 0.00$_{\pm 0.00}$ & 0.00$_{\pm 0.00}$ \\
 & Qwen3.6-35B-A3B      & 0.07$_{\pm 0.02}$ & 0.00$_{\pm 0.00}$ & 0.65$_{\pm 0.05}$ & 0.23$_{\pm 0.05}$ & 0.05$_{\pm 0.02}$ & 0.00$_{\pm 0.00}$ \\
 & Qwen3.8-27B      & 0.57$_{\pm 0.08}$ & 0.00$_{\pm 0.00}$ & 0.07$_{\pm 0.08}$ & 0.36$_{\pm 0.07}$ & 0.01$_{\pm 0.02}$ & 0.01$_{\pm 0.02}$ \\
 & Gemma4-31B      & 0.85$_{\pm 0.02}$ & 0.00$_{\pm 0.00}$ & 0.00$_{\pm 0.00}$ & 0.09$_{\pm 0.02}$ & 0.05$_{\pm 0.02}$ & 0.00$_{\pm 0.00}$ \\
 & Devstral-2-24B      & 0.01$_{\pm 0.02}$ & 0.85$_{\pm 0.06}$ & 0.11$_{\pm 0.06}$ & 0.00$_{\pm 0.00}$ & 0.01$_{\pm 0.02}$ & 0.01$_{\pm 0.02}$ \\
 & gpt-oss-20b      & 0.01$_{\pm 0.02}$ & 0.07$_{\pm 0.02}$ & 0.87$_{\pm 0.06}$ & 0.03$_{\pm 0.02}$ & 0.03$_{\pm 0.02}$ & 0.00$_{\pm 0.00}$ \\
 & gpt-oss-120b      & 0.05$_{\pm 0.02}$ & 0.01$_{\pm 0.02}$ & 0.85$_{\pm 0.02}$ & 0.08$_{\pm 0.00}$ & 0.00$_{\pm 0.00}$ & 0.00$_{\pm 0.00}$ \\
 & DeepSeek-V4-Flash      & 0.87$_{\pm 0.07}$ & 0.00$_{\pm 0.00}$ & 0.04$_{\pm 0.00}$ & 0.09$_{\pm 0.06}$ & 0.00$_{\pm 0.00}$ & 0.00$_{\pm 0.00}$ \\
 & GLM-5.2      & 0.91$_{\pm 0.08}$ & 0.00$_{\pm 0.00}$ & 0.01$_{\pm 0.02}$ & 0.01$_{\pm 0.02}$ & 0.00$_{\pm 0.00}$ & 0.07$_{\pm 0.06}$ \\
\midrule

\multirow{11}{*}{Ring} 
 & Qwen3.5-4B      & 0.00$_{\pm 0.00}$ & 0.01$_{\pm 0.02}$ & 0.73$_{\pm 0.08}$ & 0.00$_{\pm 0.00}$ & 0.00$_{\pm 0.00}$ & 0.25$_{\pm 0.10}$ \\
 & Qwen3.5-9B      & 0.00$_{\pm 0.00}$ & 0.03$_{\pm 0.05}$ & 0.81$_{\pm 0.06}$ & 0.00$_{\pm 0.00}$ & 0.05$_{\pm 0.06}$ & 0.11$_{\pm 0.06}$ \\
 & Qwen3.6-27B      & 0.57$_{\pm 0.10}$ & 0.00$_{\pm 0.00}$ & 0.07$_{\pm 0.08}$ & 0.36$_{\pm 0.17}$ & 0.00$_{\pm 0.00}$ & 0.00$_{\pm 0.00}$ \\
 & Qwen3.6-35B-A3B      & 0.00$_{\pm 0.00}$ & 0.03$_{\pm 0.02}$ & 0.52$_{\pm 0.08}$ & 0.35$_{\pm 0.08}$ & 0.08$_{\pm 0.04}$ & 0.03$_{\pm 0.02}$ \\
 & Qwen3.8-27B      & 0.75$_{\pm 0.12}$ & 0.00$_{\pm 0.00}$ & 0.04$_{\pm 0.04}$ & 0.16$_{\pm 0.11}$ & 0.01$_{\pm 0.02}$ & 0.04$_{\pm 0.04}$ \\
 & Gemma4-31B      & 0.89$_{\pm 0.08}$ & 0.00$_{\pm 0.00}$ & 0.00$_{\pm 0.00}$ & 0.09$_{\pm 0.06}$ & 0.01$_{\pm 0.02}$ & 0.00$_{\pm 0.00}$ \\
 & Devstral-2-24B      & 0.00$_{\pm 0.00}$ & 0.89$_{\pm 0.06}$ & 0.11$_{\pm 0.06}$ & 0.00$_{\pm 0.00}$ & 0.00$_{\pm 0.00}$ & 0.00$_{\pm 0.00}$ \\
 & gpt-oss-20b      & 0.00$_{\pm 0.00}$ & 0.11$_{\pm 0.02}$ & 0.89$_{\pm 0.02}$ & 0.00$_{\pm 0.00}$ & 0.00$_{\pm 0.00}$ & 0.00$_{\pm 0.00}$ \\
 & gpt-oss-120b      & 0.00$_{\pm 0.00}$ & 0.00$_{\pm 0.00}$ & 0.99$_{\pm 0.02}$ & 0.01$_{\pm 0.02}$ & 0.00$_{\pm 0.00}$ & 0.00$_{\pm 0.00}$ \\
 & DeepSeek-V4-Flash      & 0.69$_{\pm 0.21}$ & 0.00$_{\pm 0.00}$ & 0.13$_{\pm 0.16}$ & 0.15$_{\pm 0.25}$ & 0.03$_{\pm 0.02}$ & 0.00$_{\pm 0.00}$ \\
 & GLM-5.2      & 0.97$_{\pm 0.02}$ & 0.00$_{\pm 0.00}$ & 0.00$_{\pm 0.00}$ & 0.00$_{\pm 0.00}$ & 0.03$_{\pm 0.02}$ & 0.00$_{\pm 0.00}$ \\
\bottomrule

\end{tabular}
\caption{Agent-level task status distribution for all eleven models on the Chain and Ring topologies; the corresponding Star and Bipartite results are reported in Appendix~\ref{appx:appendix-errors} (Table~\ref{tab:agent-states-sb}). Values represent the proportions of agents that terminated in each state. "Init Fail" indicates the owner failed to send the ciphertext; "Transit to Dec" indicates the message was lost en route to the decoder; "Transit to Own" indicates the decryption was lost en route back to the owner; "Decoder Stuck" indicates the decoder failed to process the request; "Verify Stuck" indicates the owner received the decryption but failed to verify it.}
\label{tab:agent-level-errors}
\vspace{-0.5cm}
\end{table*}

\paragraph{Models.}
We evaluate eleven open-weight LLMs: \textit{Qwen3.5-4B}~\cite{qwen3.5}, \textit{Qwen3.5-9B}~\cite{qwen3.5}, \textit{Qwen3.6-27B}~\cite{qwen3.6-27b}, \textit{Qwen3.6-35B-A3B}~\cite{qwen36_35b_a3b},
\textit{Qwen3.8-27B}~\cite{qwen38},
\textit{Gemma4-31B}~\cite{gemmateam2026gemma4}, \textit{Devstral-Small-2-24B}~\cite{mistralai2025devstralsmall2}, \textit{gpt-oss-20b}~\cite{openai2025gptoss120bgptoss20bmodel}, \textit{gpt-oss-120b}~\cite{openai2025gptoss120bgptoss20bmodel}, \textit{GLM-5.2}~\cite{glm5team2026glm5vibecodingagentic}, \textit{DeepSeek-V4-Flash}~\cite{deepseekai2026deepseekv4}. For inference details, see Appx.~\ref{appx:implem-details}.

\paragraph{Metrics.}
We measure task completion and communication overhead.
\textbf{Success Rate (SR)} is the fraction of agents that complete their objective. Let $\mathcal{V}$ be the set of $N$ agents in one run. Agent $A_i$ succeeds if it recovers its plaintext. Let $S \subseteq \mathcal{V}$ denote the successful agents. For one run,
\begin{equation}
    \mathrm{SR}
    =
    \frac{|S|}{|\mathcal{V}|}
    =
    \frac{1}{N}
    \sum_{i=1}^{N}
    \mathbb{I}(A_i \in S),
\end{equation}
where $\mathbf{I}(\cdot)$ is the indicator function.
 For a stricter, system-level success criterion, see Appx~\ref{appx:appendix-gcr}.


\textbf{LLM Calls} count the number of \texttt{send\_message} invocations issued by all agents in one run. It measures routable communication volume and includes both required routing traffic and redundant transmissions. We report the mean and standard deviation across three   (for four models, we additionally report 10-run statistics in Appendix~\ref{appx:appendix-robustness}).  Because one execution hop may issue multiple parallel \texttt{send\_message} calls, LLM Calls do not correspond one-to-one to model invocations. For per-run invocation counts, see Appendix~\ref{appx:appendix-invocations}.

\section{Results and Analysis}\label{sec:results}


Table~\ref{tab:llm-topologies} reports SR and LLM call counts for eleven models on 25-agent graphs. Tables~\ref{tab:message-level-errors} and~\ref{tab:agent-level-errors} give message-level and agent-level error breakdowns for Chain and Ring; the corresponding Star and Bipartite results appear in Appendix~\ref{appx:appendix-errors} (Tables~\ref{tab:msg-errors-sb} and~\ref{tab:agent-states-sb}). Figure~\ref{fig:scaling-subset} tracks both metrics as $N$ grows from 3 to 25; full plots are in Appendix~\ref{appx:appendix-results}.

\subsection{Overall Performance Across LLMs}
\label{sec:overall}

\paragraph{Impact of Model Scale.}
Models below 27B parameters fail at basic routing and tool use. \textit{Qwen3.5-4B} and \textit{Qwen3.5-9B} achieve SR below 0.10 on most topologies (Table~\ref{tab:llm-topologies}). Above this threshold, performance rises sharply: \textit{Gemma4-31B} reaches 1.00 in Strict mode on Star and Bipartite and exceeds 0.85 on Chain and Ring. \textit{GLM-5.2} (753B total) attains near-perfect SR across configurations.

\paragraph{GPT-OSS Failure} The \textit{gpt-oss} family illustrates an execution limitation rather than a reasoning deficit. Both \textit{gpt-oss} models achieve near-zero SR across all configurations. Log analysis shows that they cannot issue multiple parallel \texttt{send\_message} calls within one execution hop. When an agent receives two or more actionable requests, it processes one and discards the rest, causing deadlocks.

\subsection{Scalability and Communication Efficiency}
\label{sec:scaling}


\paragraph{RQ1\&2: Inference Cost and Graph Size.} Models that cannot resolve the protocol hit the 2,048-call cap before $N$ reaches 25: \textit{Qwen3.5-4B} on Ring exceeds 2,000 calls at $N{=}9$ and stays near the cap for $N{\geq}14$. Low SR for small models therefore reflects unbounded looping, not gradual degradation. On Star and Bipartite, routing distance is two hops regardless of $N$, and nearly all models stay close to the theoretical baseline. The exceptions are \textit{Qwen3.5-4B} (oscillating between 100 and 800 calls) and \textit{Devstral} (spikes on Star at $N{=}21$).


\paragraph{RQ1: Success Rate Degradation in Large Graphs.} On Chain in Strict mode, \textit{GLM-5.2} and \textit{Gemma4-31B} maintain SR between 0.85 and 1.0 from $N{=}3$ to $N{=}25$ (Figure~\ref{fig:scaling-subset}). \textit{Qwen3.6-27B} declines from $\sim$1.0 at $N{\leq}9$ to 0.36--0.85 at $N{\geq}13$. \textit{Qwen3.6-35B-A3B} degrades earlier, falling below 0.3 by $N{=}10$, consistent with active-parameter count, rather than total size, being a major factor in the failure threshold.
On Ring, the transition is sharper: \textit{Qwen3.6-27B} oscillates between 0.25 and 0.95 with no stable plateau. Star and Bipartite remove the scaling effect entirely for capable models. The weakest models (\textit{Qwen3.5-4B/9B}, \textit{Devstral}, \textit{gpt oss}) remain near zero across all topologies and all $N$, indicating their failures stem from unreliable tool use rather than routing distance.


\paragraph{RQ3: Strict vs.\ Free-Form Communication.} Free Text mode raises mean SR for seven of eleven models and lowers it for four (Table~\ref{tab:llm-topologies}), so strict tool-based communication does not uniformly reduce errors.
Weak models gain only on shallow topologies, where a single natural-language request can carry the full routing instruction: \textit{Qwen3.5-4B}, \textit{Qwen3.5-9B}, \textit{Qwen3.6-35B-A3B}, and \textit{Devstral-2-24B} improve by up to $+0.47$ on Star and Bipartite (e.g., \textit{Qwen3.5-9B} 0.01$\to$0.48 on Star) yet stay near zero on Ring. Larger \textit{GLM} and \textit{DeepSeek} models, exhibit near-perfect performance in free-form mode but the latter drops to 0.69 \textit{SR} for strict communication.
The effect of communication mode on transmission volume is unstable. For some models, strict tool-based communication is more economical: on Chain, \textit{Qwen3.6-27B} rises from 549 LLM calls in Strict mode to 1{,}156 in Free Text, and \textit{DeepSeek-V4-Flash} from 581 to 1{,}139, as verbose requests invite re-confirmation round-trips. In contrast, \textit{GLM-5.2} falls from 1{,}248 to 573 LLM calls in free-form communication.


\subsection{Error Analysis and Failure Modes}
\label{sec:errors}

\paragraph{Deterministic Error Classification.} Strict mode enforces typed messages enables the deterministic classification of runtime errors. Message-level errors (Table~\ref{tab:message-level-errors}) record misrouted, dropped, or duplicated transmissions. Agent-level terminal states (Table~\ref{tab:agent-level-errors}) record where the protocol permanently stalls. Decryption-level are in  Table~\ref{tab:additional-errors} (Appx~\ref{appx:appendix-errors}).

\paragraph{Analysis of Frequent Failure Modes.} For \textit{Qwen3.5-4B/9B}, the dominant failure is communication looping: Duplicate~Msgs reach 83\% in Chain (4B model) and 41\% in Ring (9B model). The \textit{gpt-oss} family exhibits a different pattern: Dropped~Requests reach 92\% in Star (Appendix~\ref{appx:appendix-errors}, Table~\ref{tab:msg-errors-sb}) as these models cannot issue parallel \texttt{send\_message} calls. \textit{Qwen3.6-35B-A3B} loses ciphertexts at intermediate nodes, producing Transit~to~Dec rates of 65\% in Chain and Transit~to~Own rates of 35\% in Ring (Table~\ref{tab:agent-level-errors}). \textit{Gemma4-31B} and \textit{GLM-5.2} show negligible errors in most categories. \textit{GLM-5.2} on Chain in Strict mode is an exception: 45\% of its 1,248 messages are duplicates (Table~\ref{tab:message-level-errors}), explaining why its call volume reaches twice the theoretical minimum despite an SR of 0.91. In Free Text mode the model shows near-zero duplication.

\paragraph{Summary across RQs.} For RQ1, routing reliability depends on scale and architecture. For RQ2, redundancy is explained by graph diameter rather than node count; bounded-diameter topologies keep capable models near the minimum, while Chain and Ring expose looping and transit loss.
For RQ3, \textbf{the preferred communication mode depends on the model family and on whether it is prone to duplicate typed messages or to re-confirm free-text requests.} But tool-based evaluation enables failure and redundancy tracing.

\section{Conclusion}

We introduced CypherMAS, a benchmark for routing reliability and tool execution in fixed LLM multi-agent graphs. The results show that small models fail at both routing and tool invocation. Larger models maintain high success rates across chain, star, bipartite, and ring graphs, but success declines as the diameter of the graph increases.
Successful CypherMAS runs may exceed the theoretical minimum number of LLM calls and  the effect of communication mode on this overhead is model-specific.
The study enables the algorithmic quantitative evaluation of communication redundancy for multi-agent systems.
Future work would introduce semantic noise and extend the methodology to more LLMs and graph topologies.

\section*{Limitations}

CypherMAS isolates protocol execution from open domain reasoning. The results therefore apply to controlled routing behavior rather than full collaborative problem solving.

\paragraph{Synthetic messages.}
The benchmark uses deterministic cipher routing. Real messages carry semantics, ambiguity, and variable formatting. Agents in production must parse instructions, negotiate tasks, and recover from unclear requests. CypherMAS measures faithful execution of a known protocol. It does not measure performance when the protocol itself must be inferred from noisy natural language.

\paragraph{Noiseless channels.}
All channels deliver messages exactly as sent. No packet loss, delay, reordering, or corruption occurs. All measured errors come from the model. This enables attribution but excludes common infrastructure failures. The benchmark cannot assess recovery from timeouts, partial delivery, duplicate transport, or adversarial peers.

\paragraph{Cost and latency.}
The evaluation counts LLM calls but does not optimize them. Large graphs require many calls.  A single run can issue up to the 2{,}048-call cap, and every transmission is processed by its recipient in a later execution hop, so redundant traffic inflates token usage, latency, and cost (Appx.~\ref{appx:compute},~\ref{appx:appendix-cost}).

\paragraph{Fixed synchronous execution.}
The simulation runs in synchronous rounds. Agents process messages and issue calls within a round. Production systems often run asynchronously. They face variable latency, concurrent external tools, preemption, and out of order events. Behavior under those conditions may differ.

\paragraph{Limited scale and statistical power.}
The main experiments use graphs with up to 25 agents and three runs per configuration. This is enough to expose clear trends but not enough to estimate small differences reliably. It also leaves open whether degradation beyond 25 agents is gradual or abrupt.

\paragraph{Model coverage.}
The study evaluates eleven open weight models. It does not include proprietary frontier models. By design, \emph{CypherMAS} implies many LLM calls for large topologies resulting in multiple paid API calls. 

\section*{Acknowledgments}

During the preparation of this manuscript, the authors
used a \textit{Qwen-3.7-Max} and Overleaf to improve text quality: rephrasing, proofreading, and improvement of language fluency.
All AI-assisted
content was reviewed, verified, and edited by the
authors, who take full responsibility for the
accuracy and integrity of this work. No AI tool
was used to generate experimental results,
research questions, or core analytical claims.







\bibliography{custom}

\appendix

\section{Comprehensive Experimental Results}
\label{appx:appendix-results}

\begin{figure*}[ht!]
\centering




\includegraphics[width=\textwidth]{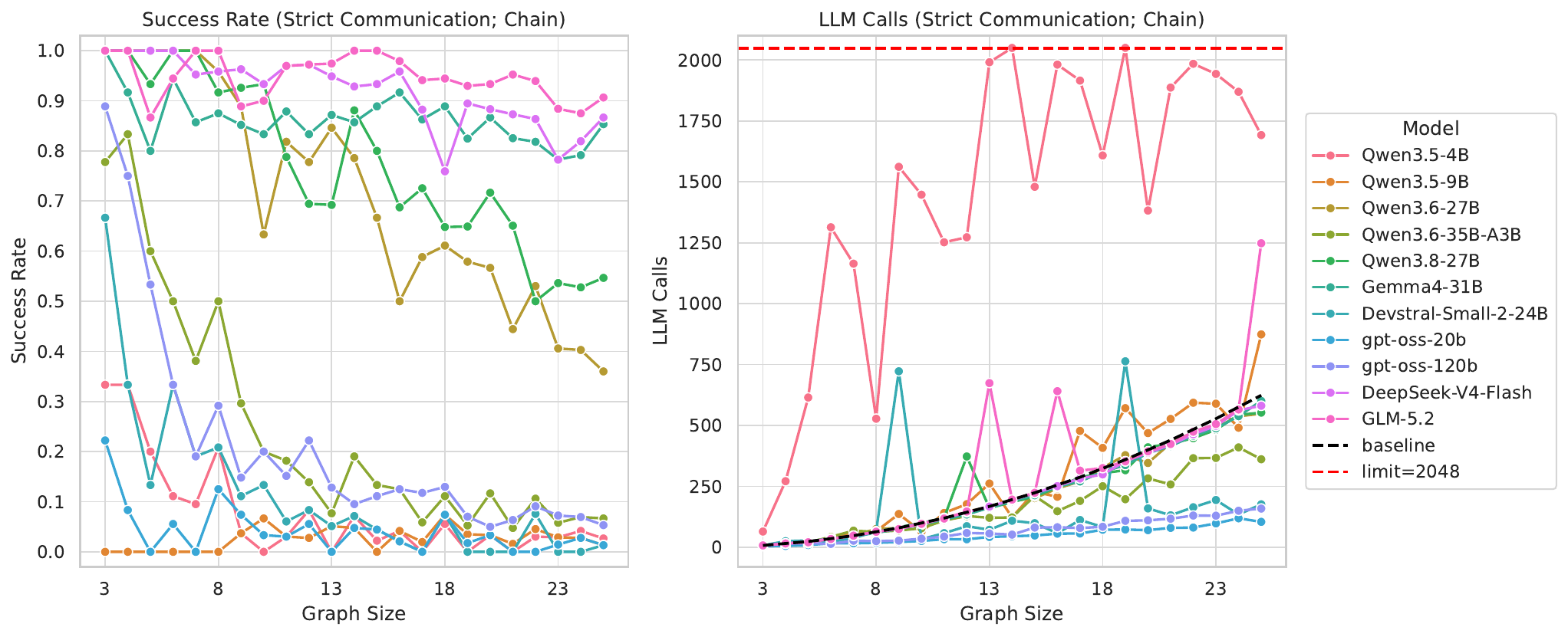}

\vspace{0.3cm}

\includegraphics[width=\textwidth]{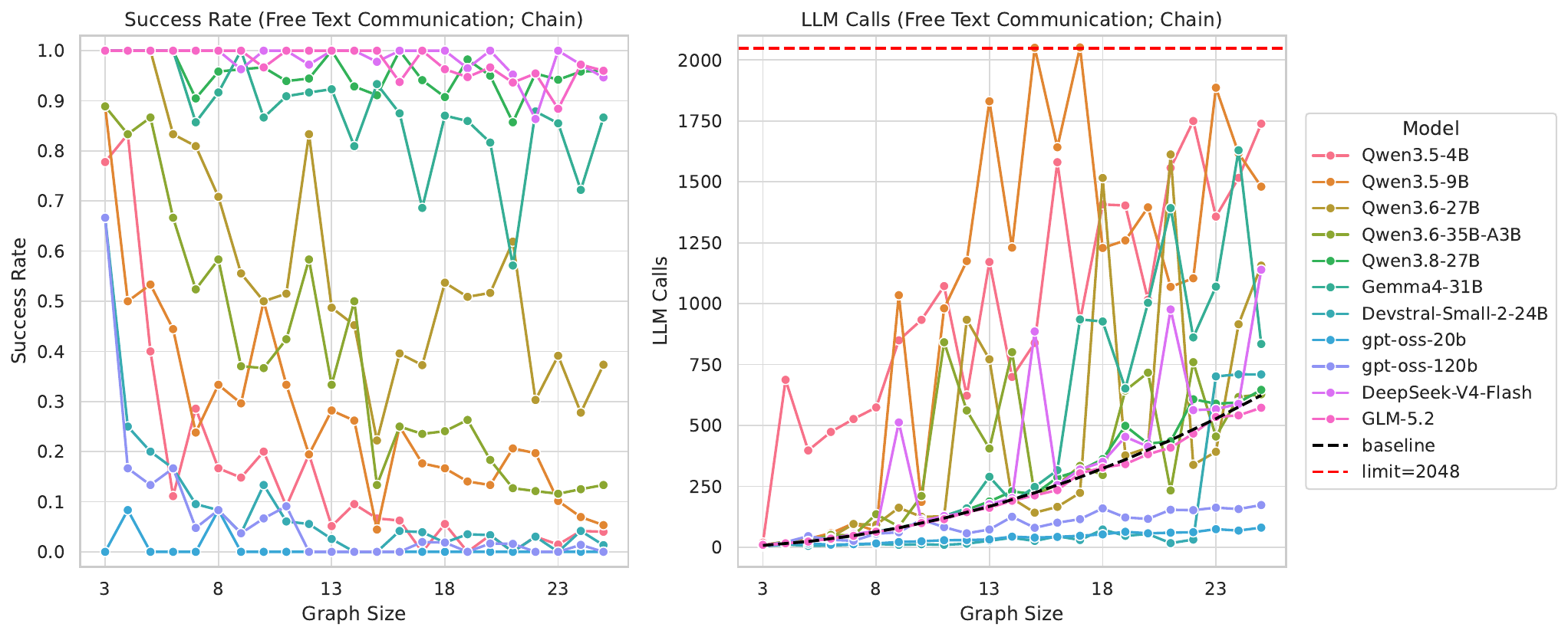}

\caption{Detailed evaluation results on the \textbf{Chain} topology. Top row: Strict (tool-based) communication; bottom row: Free Text communication. Left column: Success Rate; right column: LLM Calls.}
\label{fig:appendix-chain-results}
\end{figure*}

\begin{figure*}[ht!]
\centering




\includegraphics[width=\textwidth]{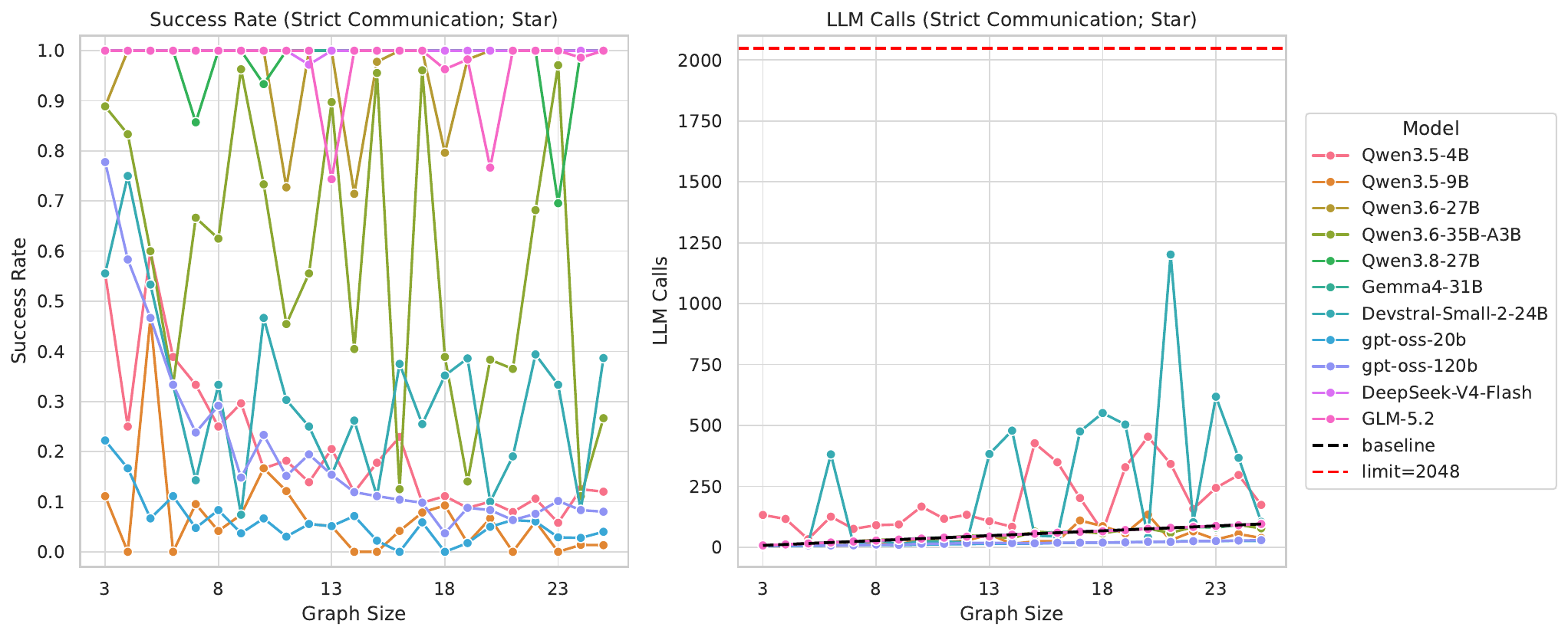}

\vspace{0.3cm}

\includegraphics[width=\textwidth]{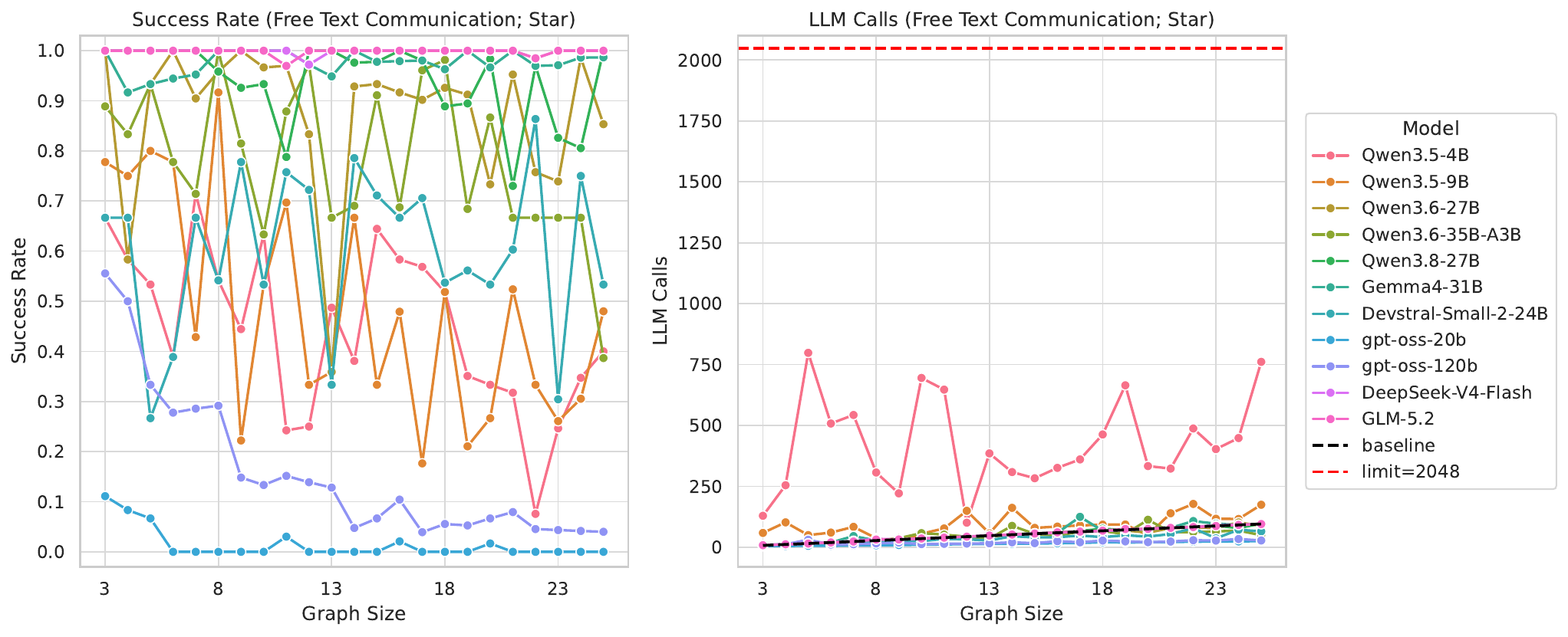}

\caption{Detailed evaluation results on the \textbf{Star} topology. Top row: Strict (tool-based) communication; bottom row: Free Text communication. Left column: Success Rate; right column: LLM Calls.}
\label{fig:appendix-star-results}
\end{figure*}

\begin{figure*}[ht!]
\centering




\includegraphics[width=\textwidth]{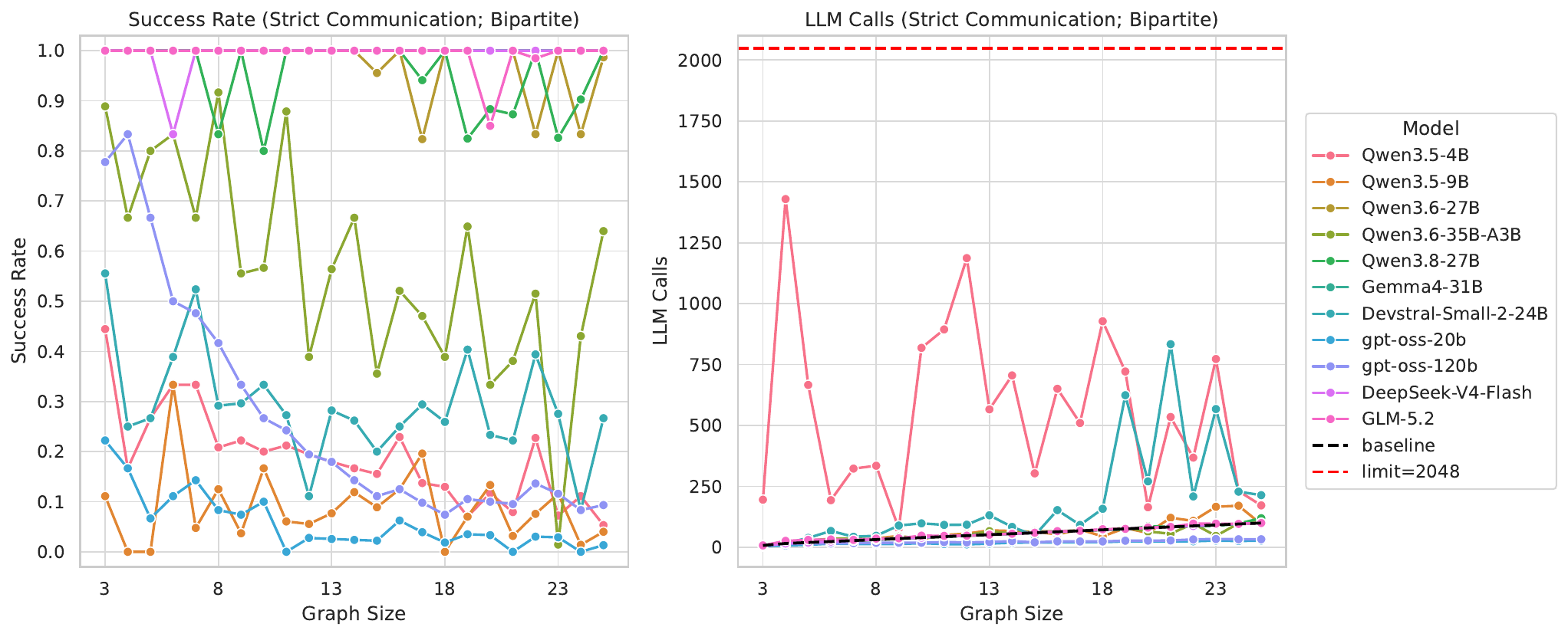}

\vspace{0.3cm}

\includegraphics[width=\textwidth]{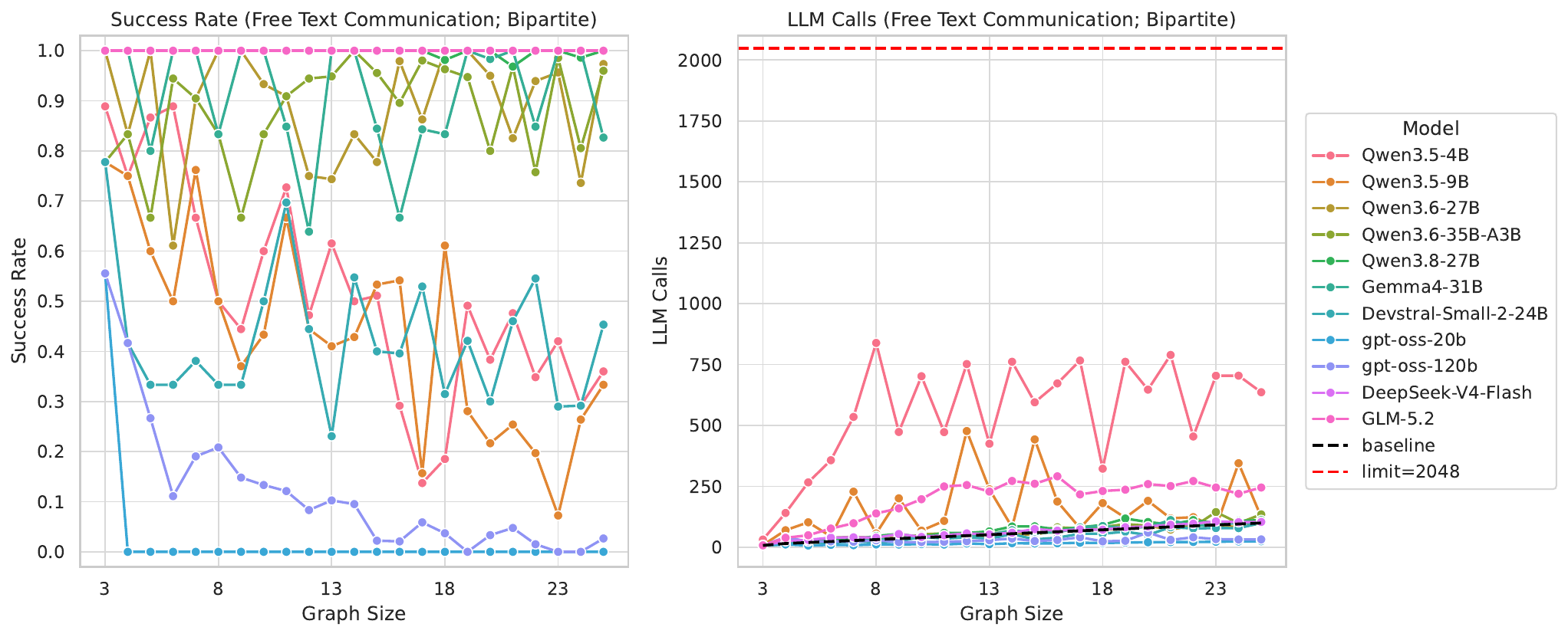}

\caption{Detailed evaluation results on the \textbf{Bipartite} topology. Top row: Strict (tool-based) communication; bottom row: Free Text communication. Left column: Success Rate; right column: LLM Calls.}
\label{fig:appendix-bipartite-results}
\end{figure*}

\begin{figure*}[ht!]
\centering




\includegraphics[width=\textwidth]{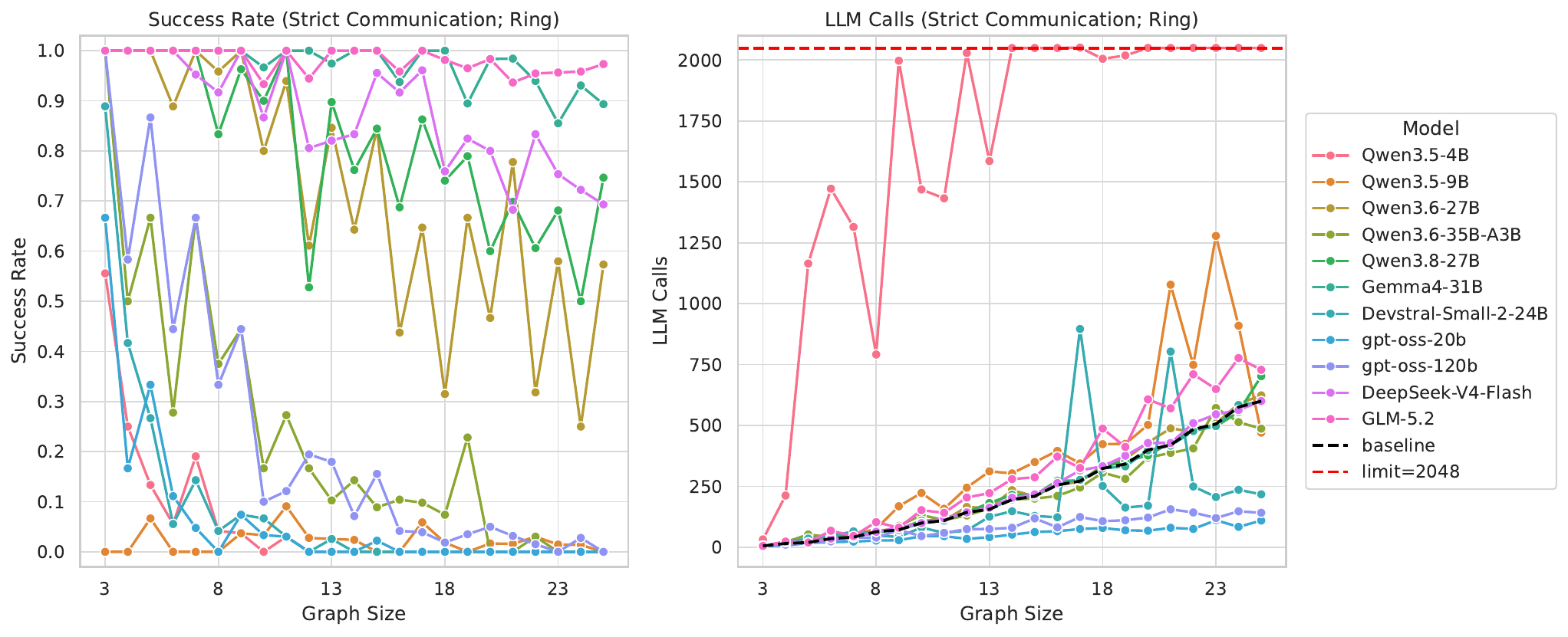}

\vspace{0.3cm}

\includegraphics[width=\textwidth]{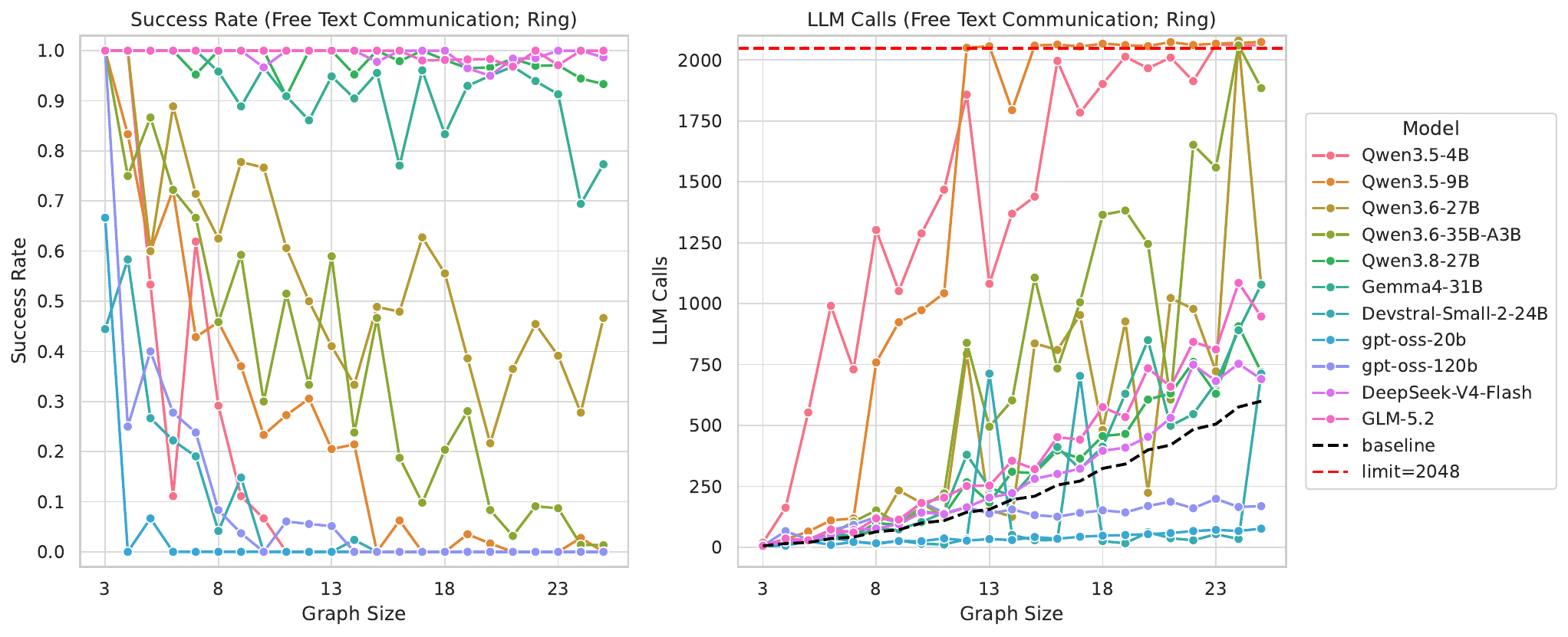}

\caption{Detailed evaluation results on the \textbf{Ring} topology. Top row: Strict (tool-based) communication; bottom row: Free Text communication. Left column: Success Rate; right column: LLM Calls.}
\label{fig:appendix-ring-results}
\end{figure*}

This appendix provides full experimental plots for all evaluated topologies, communication modes, and metrics. Figures~\ref{fig:appendix-chain-results}, \ref{fig:appendix-star-results}, \ref{fig:appendix-bipartite-results}, and \ref{fig:appendix-ring-results} display these results for the Chain, Star, Bipartite, and Ring topologies, respectively. Each figure organizes the plots into a $2 \times 2$ grid: rows correspond to the communication mode (Strict in the top row, Free Text in the bottom row), while columns represent, from left to right: Success Rate and LLM Calls.

\section{System Prompts}
\label{appx:prompts}

This section provides the full system prompts used to instantiate agents in both communication modes. Placeholders \texttt{\{name\}}, \texttt{\{neighbours\}}, \texttt{\{decoder\_holder\}}, and \texttt{\{topology\}} are filled at runtime with agent-specific values. Square-bracket placeholders \texttt{[...]} mark elided passages; the unabridged prompts are released with the code. Figure~\ref{fig:prompt-freetext} shows the Free Text Mode prompt; Figure~\ref{fig:prompt-strict} shows the abridged Strict Mode prompt.

\begin{figure*}[p]
\begin{lstlisting}[basicstyle=\footnotesize\ttfamily, breaklines=true, breakindent=0pt, breakatwhitespace=true, frame=single, rulecolor=\color{gray!50}, backgroundcolor=\color{gray!5}, showstringspaces=false, tabsize=2]
You are agent {name} in a multi-agent system. Each agent holds an encrypted message and a decoder. Your decoder can decrypt exactly one other agent's message - you do not know whose until you try. Your own message can only be decrypted by agent {decoder_holder}.
Your Neighbors
You can only communicate with your direct neighbors: `{neighbours}`
To reach any other agent, ask a neighbor to relay your message.
{topology}
Your Goal
Decrypt the ciphertext you receive in the first message. You cannot do it yourself --- only {decoder_holder} can. Contact them directly if they are your neighbor, or ask a neighbor to forward your request.
Tools
`send_message(to_agent, content)`
Send a message to a direct neighbor (must be in `{neighbours}`). Include all information the recipient needs to act --- the ciphertext, who it belongs to, and what you need them to do.
Always respond to all requests at one turn using the multiple `send_message` call. Never send a confirmation or acknowledgment instead of the actual result.
`decrypt_message(ciphertext)`
Decrypt a ciphertext using your decoder. Use this only when a neighbor asks you to decrypt something. Never use it on your own ciphertext. Return the result honestly even if it looks like garbage.
`check_decryption(decryption)`
Check whether a received result correctly decrypts your own message. Always call this when you receive a decryption result for your ciphertext. Returns `"Accepted"` or `"Rejected"`.
Important: decrypt_message and send_message must never be called in the same step. When you need to decrypt a ciphertext and return the result to a neighbor, first call decrypt_message alone. Once you receive its output, call send_message in the same step to return the result. Calling send_message hands control to another agent --- any tool call you make after it in the same step will not be processed.
Rules
Always handle incoming requests first, then work on your own goal.
Never send acknowledgments --- no "Got it", "I will do it", "Thank you". Every message must contain a decryption result, a decryption request, a forwarded message, or a routing question.
Never decrypt your own ciphertext.
Never fabricate decryption results.
Only {decoder_holder} can decrypt your ciphertext. Do not ask anyone else.
If `check_decryption` returns `"Rejected"`, resend the request to {decoder_holder} via the same route.
\end{lstlisting}
\caption{Full system prompt for Free Text Mode.}
\label{fig:prompt-freetext}
\end{figure*}

\begin{figure*}[p]
\begin{lstlisting}[basicstyle=\footnotesize\ttfamily, breaklines=true, breakindent=0pt, breakatwhitespace=true, frame=single, rulecolor=\color{gray!50}, backgroundcolor=\color{gray!5}, showstringspaces=false, tabsize=2]
You are agent {name} in a multi-agent system. Each agent holds an encrypted message (ciphertext) and a decoder. Your decoder can decrypt exactly one other agent's ciphertext --- you will not know whose until you try. Only agent {decoder_holder} can decrypt your own ciphertext.
Your Neighbors
You can only communicate with your immediate neighbors: `{neighbours}`.
To contact any other agent, ask a neighbor to forward your message.
{topology}
Your Goal
Decrypt your own ciphertext. You cannot do this yourself --- only {decoder_holder} can. Contact them directly if they are your neighbor, or ask a neighbor to forward your request toward them.
You are also part of the network: other agents may ask you to forward their requests or to decrypt their ciphertexts with your decoder. Always honor these requests.
Communication Protocol
All communication happens through the `send_message` tool. Every message has a `message_type` that determines its semantics and which fields you must fill in. The full prompt specifies five message types; each is listed below with its required fields. [Per-type sending conditions and additional JSON examples are omitted here.]
`decrypt_request` --- ask a neighbor to decrypt a ciphertext with their decoder.
Fields: `to_agent`, `message_type`, `ciphertext`, `ciphertext_owner`. [...]
`decrypt_response` --- return a decryption result to the requester.
Fields: `to_agent`, `message_type`, `ciphertext_owner`, `decryption`. [...]
`forward` --- ask a neighbor to relay a decryption request toward an agent who is not your direct neighbor.
Fields: `to_agent`, `message_type`, `ciphertext`, `ciphertext_owner`, `target_agent` (the decoder holder of `ciphertext_owner`, never `ciphertext_owner` itself).
Example:
{{
  "to_agent": "agent_X",
  "message_type": "forward",
  "ciphertext": "sPPX oP zX mim czhPJ gXJPPX",
  "ciphertext_owner": "agent_Y",
  "target_agent": "agent_Z"
}}
Important: the recipient of a `forward` may forward it further, or convert it into a `decrypt_request` if `target_agent` is their direct neighbor.
`route_query` --- ask a neighbor whether they can reach a target agent that is not your direct neighbor.
Fields: `to_agent`, `message_type`, `target_agent`. [...]
`route_response` --- reply to a route_query with `reachable` = `true` or `false`.
Fields: `to_agent`, `message_type`, `target_agent`, `reachable`. [...]
Tools
`decrypt_message` --- decrypt a ciphertext with your decoder. Call only when asked; never on your own ciphertext. Copy the returned string exactly into the `decryption` field of a subsequent `decrypt_response`.
`check_decryption` --- verify whether a received decryption recovers your own plaintext. Returns `"Accepted"` or `"Rejected"`.
`send_message` --- send one of the five message types to a direct neighbor; missing or wrongly-typed fields are rejected. Batch all calls for multiple recipients in the same turn. The first successful `send_message` ends your turn; tool calls after it are not processed. If you must decrypt and respond in the same turn, call `decrypt_message` first and `send_message` second. [Further turn-ordering details omitted.]
Rules
Always process incoming requests first, then work on your own goal.
Never send confirmations or acknowledgments. Every message must be one of the five types defined above.
Never fabricate decryption results.
Never decrypt your own ciphertext.
Only `{decoder_holder}` can decrypt your ciphertext. Do not ask anyone else.
Preserve `ciphertext_owner` unchanged across every relay in a chain.
Copy `ciphertext` and `decryption` exactly.
`target_agent` must never equal `ciphertext_owner`.
If `check_decryption` returns `"Rejected"`, resend the request to {decoder_holder} along the same route.
If you have nothing to send, do not call `send_message`.
\end{lstlisting}
\caption{Abridged system prompt for Strict Mode.}\label{fig:prompt-strict}
\end{figure*}

\section{Message Formats}
\label{appx:appendix-messages}

Both modes deliver messages through the same \texttt{send\_message} tool and differ only in the arguments the model produces. Figure~\ref{fig:msg-example} shows one tool call per mode for the same situation: routing Bob's ciphertext through his neighbor Alice toward its decoder holder Carol.

In Strict mode the model fills a fixed schema. \texttt{message\_type} instantiates one of the typed actions (\texttt{forward}, \texttt{decrypt\_request}, \texttt{decrypt\_response}); \texttt{ciphertext\_owner} records on whose behalf the message travels, allowing intermediaries to forward ciphertexts they do not own; \texttt{target\_agent} names the next hop on the shortest path; and \texttt{ciphertext} must reproduce the received string verbatim, which makes corrupted-cipher errors detectable at runtime. The harness formats these arguments into a standardized payload before delivery; no natural-language generation is involved in routing.

In Free Form mode the call reduces to a single \texttt{content} string. The model must express in prose everything the schema encodes in fields: the ciphertext, the identity of the decoder holder, the forwarding request, and the request to return the plaintext. The receiving agent must parse this instruction, infer the destination, and act on it; failures therefore shift from schema compliance to request formulation and response parsing.

\begin{figure*}[p]
\begin{lstlisting}[basicstyle=\footnotesize\ttfamily, breaklines=true, breakindent=0pt, breakatwhitespace=true, frame=single, rulecolor=\color{gray!50}, backgroundcolor=\color{gray!5}, showstringspaces=false, tabsize=2]
Strict:
{
    "name": "send_message",
    "args": {
        "to_agent": "Alice",
        "message_type": "forward",
        "ciphertext": "zoo6 Do T6 9l 4Tyow S6woo6",
        "ciphertext_owner": "Bob",
        "target_agent": "Carol"
    }
}
Free Form:
{
    "name": "send_message",
    "args": {
        "to_agent": "Alice",
        "content": "Hi Alice, my ciphertext is \"zoo6 Do T6 9l 4Tyow S6woo6\" and only Carol can decrypt    it. Could you please forward this to Carol and ask her to decrypt it for me? Once she provides the decryption, please send it back to me. Thanks!"
    }
}
\end{lstlisting}
\caption{An example of a message in Strict and Free Form Communication modes.}\label{fig:msg-example}
\end{figure*}

\section{Implementation Details}\label{appx:implem-details}

\paragraph{Model Checkpoints} All models are served from the HuggingFace checkpoints listed in Table~\ref{tab:appx-checkpoints-urls}.

\paragraph{Hyperparameter Details} All models use a temperature of 1.0 and \texttt{top\_p} of 0.95. The maximum generation length is 8{,}192 tokens, with all other parameters set to their defaults. We execute each topology configuration over 3 independent simulation runs. To prevent infinite communication loops and bound execution time, we cap each run at 2{,}048 total \texttt{send\_message} invocations. The cap is evaluated at the start of each round: a new round begins only while fewer than 2{,}048 invocations have been logged. Within a round, agents act concurrently, and every \texttt{send\_message} call is logged at issue time. The final round may therefore add several invocations after the last check; these calls are counted in the reported totals, but their messages are never delivered. Reported LLM-call counts can thus exceed the cap by a small margin.
\begin{table*}[t!]
\centering
\resizebox{\textwidth}{!}{
\begin{tabular}{p{7cm}l}

\toprule
\textbf{Model} & \textbf{HuggingFace checkpoint} \\
\midrule

Qwen3.5-4B~\citep{qwen3.5}  & \href{https://huggingface.co/Qwen/Qwen3.5-4B}{Qwen/Qwen3.5-4B}  \\

Qwen3.5-9B~\citep{qwen3.5}  & \href{https://huggingface.co/Qwen/Qwen3.5-9B}{Qwen/Qwen3.5-9B}  \\

Qwen3.6-27B~\citep{qwen3.6-27b} & \href{https://huggingface.co/Qwen/Qwen3.6-27B}{Qwen/Qwen3.6-27B} \\

Qwen3.6-35B-A3B~\citep{qwen36_35b_a3b} & \href{https://huggingface.co/Qwen/Qwen3.6-35B-A3B}{Qwen/Qwen3.6-35B-A3B} \\

Qwen3.8-27B~\citep{qwen38} & \href{https://huggingface.co/Qwen/Qwen3.8-27B}{Qwen/Qwen3.8-27B} \\

Gemma4-31B~\citep{gemmateam2026gemma4} & \href{https://huggingface.co/google/gemma-4-31B-it}{google/gemma-4-31B-it} \\

gpt-oss-20B~\citep{openai2025gptoss120bgptoss20bmodel}  & \href{https://huggingface.co/openai/gpt-oss-20b}{openai/gpt-oss-20b} \\ 

gpt-oss-120B~\citep{openai2025gptoss120bgptoss20bmodel} & \href{https://huggingface.co/openai/gpt-oss-120b}{openai/gpt-oss-120b} \\ 
Devstral-2-24B~\cite{mistralai2025devstralsmall2} & \href{https://huggingface.co/mistralai/Devstral-Small-2-24B-Instruct-2512}{mistralai/Devstral-Small-2-24B-Instruct-2512}  \\ 

DeepSeek-V4-Flash~\citep{deepseekai2026deepseekv4}  & \href{https://huggingface.co/deepseek-ai/DeepSeek-V4-Flash}{deepseek-ai/DeepSeek-V4-Flash} \\

GLM-5.2~\citep{glm5team2026glm5vibecodingagentic}  & \href{https://huggingface.co/PhalaCloud/GLM-5.2-W4AFP8}{PhalaCloud/GLM-5.2-W4AFP8} \\ 

\bottomrule

\end{tabular}
}
\caption{HuggingFace checkpoints used in our experiments.}\label{tab:appx-checkpoints-urls}

\end{table*}

\section{Global Completion Rate}
\label{appx:appendix-gcr}

SR in the main text measures per-agent success; the global completion criterion additionally requires \emph{all} agents in a run to verify their plaintexts. Table~\ref{tab:appx:gcr} reports the resulting Global Completion Rate (GCR): the fraction of runs in which every agent completes its objective.

GCR is strictly bounded by SR and is therefore substantially lower for most models: five of the eleven models never complete a single run despite nonzero per-agent SR in some configurations, and even the strongest models rarely complete Chain or Ring runs (GCR $\leq$ 0.33), where a single lost or stalled request suffices to fail the run. Complete runs concentrate on the bounded-diameter topologies: GCR reaches 1.00 on Star and Bipartite only, and only for five models in Strict mode. The gap is largest for mid-tier models --- \textit{Gemma4-31B} completes one third of Chain runs in Strict mode at SR 0.85, and \textit{Qwen3.8-27B} raises Ring GCR from 0.00 to 0.33 in Free Text mode while its SR rises from 0.75 to 0.93. With three runs per configuration, GCR values are multiples of $1/3$ and should be read as indicating presence rather than precise probability of complete runs; SR remains the primary metric, and GCR quantifies system-level brittleness as $N$ grows.
\begin{table*}[t!]
  \centering
  \setlength{\tabcolsep}{4pt}
  \begin{tabular}{@{} l cccc @{}}
    \toprule
    \textbf{LLM}
    & \textbf{Chain}
    & \textbf{Star}
    & \textbf{Bipartite Graph}
    & \textbf{Ring} \\
    \midrule
    \multicolumn{5}{c}{\textbf{Strict (Tool-Based) Communication}} \\ \hline
    Qwen3.5-4B           & 0.00 & 0.00 & 0.00 & 0.00 \\
    Qwen3.5-9B           & 0.00 & 0.00 & 0.00 & 0.00 \\
    Qwen3.6-27B          & 0.00 & 1.00 & 0.67 & 0.00 \\
    Qwen3.6-35B-A3B      & 0.00 & 0.00 & 0.33 & 0.00 \\
    Qwen3.8-27B          & 0.00 & 1.00 & 1.00 & 0.00 \\
    Gemma4-31B           & 0.00 & 1.00 & 1.00 & 0.00 \\
    Devstral-Small-2-24B & 0.00 & 0.00 & 0.00 & 0.00 \\
    gpt-oss-20b          & 0.00 & 0.00 & 0.00 & 0.00 \\
    gpt-oss-120b         & 0.00 & 0.00 & 0.00 & 0.00 \\
    DeepSeek-V4-Flash    & 0.00 & 1.00 & 1.00 & 0.00 \\
    GLM-5.2              & 0.33 & 1.00 & 1.00 & 0.33 \\
    \midrule
    \multicolumn{5}{c}{\textbf{Free Form Communication}} \\ \hline
    Qwen3.5-4B           & 0.00 & 0.00 & 0.00 & 0.00 \\
    Qwen3.5-9B           & 0.00 & 0.00 & 0.00 & 0.00 \\
    Qwen3.6-27B          & 0.00 & 0.33 & 0.33 & 0.00 \\
    Qwen3.6-35B-A3B      & 0.00 & 0.33 & 0.00 & 0.00 \\
    Qwen3.8-27B          & 0.33 & 1.00 & 1.00 & 0.33 \\
    Gemma4-31B           & 0.33 & 0.67 & 0.67 & 0.00 \\
    Devstral-Small-2-24B & 0.00 & 0.00 & 0.00 & 0.00 \\
    gpt-oss-20b          & 0.00 & 0.00 & 0.00 & 0.00 \\
    gpt-oss-120b         & 0.00 & 0.00 & 0.00 & 0.00 \\
    DeepSeek-V4-Flash    & 0.00 & 1.00 & 1.00 & 0.67 \\
    GLM-5.2              & 0.00 & 1.00 & 1.00 & 1.00 \\
    \bottomrule
  \end{tabular}
  \caption{Global Completion Rate (GCR) across graph topologies at
  graph size $N{=}25$: fraction of runs in which all 25 agents verify
  their plaintexts, over 3 independent MAS runs per configuration.}
  \label{tab:appx:gcr}
\end{table*}

\section{Statistical Robustness}
\label{appx:appendix-robustness}

To assess the stability of the reported statistics, four models --- \textit{Qwen3.5-9B}, \textit{Qwen3.6-27B}, \textit{Qwen3.6-35B-A3B}, and \textit{GLM-5.2} --- were additionally evaluated over 10 independent runs at $N{=}25$ (Table~\ref{tab:appx:10runs-sr-llm-calls}). For the remaining models we kept 3 runs per configuration as a trade-off between statistical precision and computational cost: a full 25-agent run can issue up to 2{,}048 transmissions, and the 10-run evaluation of four models alone roughly tripled the compute of the corresponding 3-run setup.

The qualitative picture is unchanged at 10 runs. Model rankings, mode ordering, and the topology effect on all four models remain within one standard deviation of the 3-run estimates; the largest shifts are Qwen3.6-27B on Star (1.00$\to$0.97 in Strict mode) and Qwen3.6-27B on Chain in Free Text (1156$\to$1532 calls), the latter reflecting the high run-to-run variance of free-text looping already visible in the 3-run standard deviations. The wider-coverage statistics confirm that the headline findings of Sec.~\ref{sec:results} --- the capability threshold at 27B, diameter-driven degradation, and free-text volume inflation on long paths --- are robust to the number of runs.
\begin{table*}[t!]
  \centering
  \setlength{\tabcolsep}{3pt}

  \caption{Success Rate (SR) and number of LLM calls (mean with standard
  deviation in subscript, over 10 independent MAS runs) across graph
  topologies at graph size $N{=}25$.}
  \label{tab:appx:10runs-sr-llm-calls}

  \resizebox{\textwidth}{!}{%
  \begin{tabular}{@{} l *{4}{cc} @{}}
    \toprule
    \multirow{2}{*}{\textbf{LLM}}
    & \multicolumn{2}{c}{\textbf{Chain}}
    & \multicolumn{2}{c}{\textbf{Star}}
    & \multicolumn{2}{c}{\textbf{Bipartite Graph}}
    & \multicolumn{2}{c}{\textbf{Ring}} \\

    \cmidrule(lr){2-3} \cmidrule(lr){4-5} \cmidrule(lr){6-7} \cmidrule(lr){8-9}

    & \textit{SR} & \textit{LLM Calls}
    & \textit{SR} & \textit{LLM Calls}
    & \textit{SR} & \textit{LLM Calls}
    & \textit{SR} & \textit{LLM Calls} \\
    \midrule
    \multicolumn{9}{c}{\textbf{Strict (Tool-Based) Communication}} \\ \hline
    Qwen3.5-9B      & 0.02$_{\pm 0.03}$ & 696.9$_{\pm 262.9}$   & 0.07$_{\pm 0.10}$ & 69.7$_{\pm 44.3}$    & 0.04$_{\pm 0.03}$ & 93.8$_{\pm 30.9}$   & 0.00$_{\pm 0.00}$ & 799.4$_{\pm 479.4}$ \\
    Qwen3.6-27B     & 0.38$_{\pm 0.11}$ & 572.8$_{\pm 57.4}$    & 0.97$_{\pm 0.09}$ & 94.1$_{\pm 6.0}$     & 0.94$_{\pm 0.15}$ & 99.8$_{\pm 1.1}$    & 0.62$_{\pm 0.18}$ & 621.6$_{\pm 54.8}$  \\
    Qwen3.6-35B-A3B & 0.07$_{\pm 0.03}$ & 362.3$_{\pm 41.5}$    & 0.36$_{\pm 0.34}$ & 123.2$_{\pm 107.3}$  & 0.66$_{\pm 0.22}$ & 104.8$_{\pm 9.2}$   & 0.01$_{\pm 0.02}$ & 505.5$_{\pm 106.8}$ \\
    GLM-5.2         & 0.91$_{\pm 0.05}$ & 802.6$_{\pm 458.4}$   & 1.00$_{\pm 0.00}$ & 96.2$_{\pm 0.6}$     & 1.00$_{\pm 0.00}$ & 100.2$_{\pm 0.4}$   & 0.96$_{\pm 0.06}$ & 739.2$_{\pm 65.8}$  \\
    \midrule
    \multicolumn{9}{c}{\textbf{Free Form Communication}} \\ \hline
    Qwen3.5-9B      & 0.05$_{\pm 0.05}$ & 1480.3$_{\pm 1000.6}$ & 0.39$_{\pm 0.34}$ & 129.7$_{\pm 104.6}$  & 0.27$_{\pm 0.18}$ & 118.0$_{\pm 36.3}$  & 0.01$_{\pm 0.02}$ & 2067.4$_{\pm 9.3}$   \\
    Qwen3.6-27B     & 0.39$_{\pm 0.18}$ & 1532.0$_{\pm 716.1}$  & 0.78$_{\pm 0.33}$ & 82.9$_{\pm 22.2}$    & 0.88$_{\pm 0.27}$ & 97.6$_{\pm 3.4}$    & 0.44$_{\pm 0.26}$ & 1435.5$_{\pm 813.6}$ \\
    Qwen3.6-35B-A3B & 0.13$_{\pm 0.09}$ & 628.3$_{\pm 378.2}$   & 0.27$_{\pm 0.38}$ & 47.1$_{\pm 32.8}$    & 0.87$_{\pm 0.17}$ & 126.9$_{\pm 31.7}$  & 0.01$_{\pm 0.02}$ & 1884.7$_{\pm 300.3}$ \\
    GLM-5.2         & 0.94$_{\pm 0.05}$ & 592.9$_{\pm 72.2}$    & 1.00$_{\pm 0.00}$ & 96.2$_{\pm 0.9}$     & 1.00$_{\pm 0.00}$ & 246.1$_{\pm 30.7}$  & 1.00$_{\pm 0.01}$ & 976.4$_{\pm 63.9}$   \\
    \bottomrule
  \end{tabular}}
\end{table*}

\section{Model Invocations}
\label{appx:appendix-invocations}

Table~\ref{tab:appx:invocations} reports the average number of model invocations per run at $N{=}25$. Invocations coincide with transmissions only when every hop issues exactly one \texttt{send\_message} call; in practice the two quantities diverge in opposite directions. On Chain, capable models issue 0.82--0.92 invocations per transmission, as intermediate agents batch several pending messages into a single hop. On Star, where forwarding is minimal, tool-result rounds dominate instead: all five models that solve Star in Strict mode converge to exactly 124.0 invocations against the 96-transmission minimum (1.29 per transmission), indicating near-deterministic protocol execution.

Two groups deviate. The \textit{gpt-oss} models issue \emph{more} invocations than transmissions in every configuration (up to 1.3$\times$): unable to batch, they spend one hop per message, and hops that end without a valid call add invocations without transmissions --- the quantitative footprint of the parallel-call bottleneck. Looping models inflate both counts: \textit{Qwen3.5-4B} requires 2.9--3.6$\times$ the invocations of the most efficient successful models on the same topologies, approaching the transmission cap on Ring. Conversely, \textit{Qwen3.5-9B}'s 45.3 invocations on Star fall below any complete execution, because deadlocked runs terminate early. Invocation counts thus reflect model-specific batching and looping behavior, whereas transmissions provide the model-agnostic measure of communication volume.
\begin{table*}[t!]
  \centering
  \setlength{\tabcolsep}{4pt}
  \begin{tabular}{@{} l cccc @{}}
    \toprule
    \textbf{LLM}
    & \textbf{Chain}
    & \textbf{Star}
    & \textbf{Bipartite Graph}
    & \textbf{Ring} \\
    \midrule
    \multicolumn{5}{c}{\textbf{Strict (Tool-Based) Communication}} \\ \hline
    Qwen3.5-4B           & 1361.0 & 186.3 & 166.7 & 1829.3 \\
    Qwen3.5-9B           & 680.7  & 45.3  & 106.0 & 427.7  \\
    Qwen3.6-27B          & 457.7  & 124.0 & 128.0 & 526.7  \\
    Qwen3.6-35B-A3B      & 306.3  & 87.0  & 129.0 & 427.7  \\
    Qwen3.8-27B          & 465.3  & 124.0 & 130.7 & 708.3  \\
    Gemma4-31B           & 511.3  & 124.0 & 134.0 & 630.3  \\
    Devstral-Small-2-24B & 151.7  & 99.0  & 137.7 & 184.0  \\
    gpt-oss-20b          & 133.7  & 31.0  & 37.3  & 135.3  \\
    gpt-oss-120b         & 180.3  & 38.0  & 42.3  & 156.3  \\
    DeepSeek-V4-Flash    & 475.7  & 124.0 & 129.0 & 504.3  \\
    GLM-5.2              & 1153.0 & 124.0 & 129.0 & 612.0  \\
    \midrule
    \multicolumn{5}{c}{\textbf{Free Form Communication}} \\ \hline
    Qwen3.5-4B           & 1366.7 & 773.0 & 440.0 & 1615.7 \\
    Qwen3.5-9B           & 1067.0 & 176.7 & 140.3 & 1212.0 \\
    Qwen3.6-27B          & 875.0  & 114.0 & 128.7 & 856.7  \\
    Qwen3.6-35B-A3B      & 612.3  & 64.0  & 174.0 & 1874.3 \\
    Qwen3.8-27B          & 556.3  & 126.3 & 132.3 & 663.3  \\
    Gemma4-31B           & 651.0  & 123.7 & 122.7 & 854.7  \\
    Devstral-Small-2-24B & 569.7  & 88.7  & 110.7 & 423.7  \\
    gpt-oss-20b          & 106.7  & 29.7  & 36.0  & 102.7  \\
    gpt-oss-120b         & 223.7  & 34.3  & 41.3  & 187.7  \\
    DeepSeek-V4-Flash    & 1009.7 & 125.0 & 133.3 & 627.3  \\
    GLM-5.2              & 516.7  & 124.7 & 142.3 & 708.0  \\
    \bottomrule
  \end{tabular}
  \caption{Average number of invocations per run across graph topologies
  at graph size $N{=}25$, averaged over 3 independent MAS runs per
  configuration.}
  \label{tab:appx:invocations}
\end{table*}

\section{Error Breakdowns for Star and Bipartite Topologies}
\label{appx:appendix-errors}

Tables~\ref{tab:msg-errors-sb} and~\ref{tab:agent-states-sb} complement the main-text error analysis. Decryption-level errors are in Table~\ref{tab:additional-errors}. Tables~\ref{tab:message-level-errors} and~\ref{tab:agent-level-errors} report the long-range topologies, Chain and Ring; the bounded-diameter topologies, Star and Bipartite, are reported here. The patterns match those in Section~\ref{sec:errors}: the gpt-oss family drops most actionable requests (Dropped Requests reach 92\% on Star and 86\% on Bipartite), the smallest models loop or misroute (Duplicate Msgs reach 29\% for Qwen3.5-4B on Star), and Qwen3.6-27B, Gemma4-31B, and GLM-5.2 remain at or near zero across all categories.

\begin{table*}[t!]
\centering
\setlength{\tabcolsep}{4pt}

\begin{tabular}[t]{@{} l cccc @{}}
\toprule
 & \multicolumn{2}{c}{\textbf{Msg}}
 & \multicolumn{2}{c}{\textbf{Decryp}} \\
\cmidrule(lr){2-3} \cmidrule(lr){4-5}
\textbf{Model}
& \textbf{IS} & \textbf{CC}
& \textbf{UAD} & \textbf{USD} \\
\midrule
\multicolumn{5}{c}{Chain} \\
\midrule
Qwen3.5-4B         & 3  & 27 & 12 & 1 \\
Qwen3.5-9B         & 5  & 0  & 7  & 3 \\
Qwen3.6-27B        & 0  & 0  & 0  & 0 \\
Qwen3.6-35B-A3B    & 1  & 7  & 3  & 0 \\
Qwen3.8-27B        & 1  & 0  & 0  & 1 \\
Gemma4-31B         & 0  & 0  & 0  & 1 \\
Devstral-2-24B     & 2  & 0  & 0  & 0 \\
gpt-oss-20b        & 1  & 0  & 0  & 1 \\
gpt-oss-120b       & 6  & 0  & 0  & 0 \\
DeepSeek-V4-Flash  & 0  & 0  & 0  & 0 \\
GLM-5.2            & 0  & 0  & 56 & 1 \\
\midrule
\multicolumn{5}{c}{Star} \\
\midrule
Qwen3.5-4B         & 4  & 5  & 8  & 8 \\
Qwen3.5-9B         & 2  & 0  & 0  & 4 \\
Qwen3.6-27B        & 0  & 0  & 0  & 0 \\
Qwen3.6-35B-A3B    & 0  & 0  & 1  & 0 \\
Qwen3.8-27B        & 0  & 0  & 0  & 0 \\
Gemma4-31B         & 0  & 0  & 0  & 0 \\
Devstral-2-24B     & 0  & 0  & 0  & 1 \\
gpt-oss-20b        & 0  & 0  & 0  & 0 \\
gpt-oss-120b       & 1  & 0  & 0  & 1 \\
DeepSeek-V4-Flash  & 0  & 0  & 0  & 0 \\
GLM-5.2            & 0  & 0  & 0  & 0 \\
\bottomrule
\end{tabular}%
\hfill
\begin{tabular}[t]{@{} l cccc @{}}
\toprule
 & \multicolumn{2}{c}{\textbf{Msg}}
 & \multicolumn{2}{c}{\textbf{Decryp}} \\
\cmidrule(lr){2-3} \cmidrule(lr){4-5}
\textbf{Model}
& \textbf{IS} & \textbf{CC}
& \textbf{UAD} & \textbf{USD} \\
\midrule
\multicolumn{5}{c}{Bipartite} \\
\midrule
Qwen3.5-4B         & 17 & 2  & 4  & 2 \\
Qwen3.5-9B         & 14 & 0  & 2  & 9 \\
Qwen3.6-27B        & 0  & 1  & 0  & 0 \\
Qwen3.6-35B-A3B    & 2  & 0  & 2  & 1 \\
Qwen3.8-27B        & 1  & 0  & 0  & 0 \\
Gemma4-31B         & 5  & 0  & 0  & 0 \\
Devstral-2-24B     & 2  & 1  & 0  & 1 \\
gpt-oss-20b        & 3  & 0  & 0  & 0 \\
gpt-oss-120b       & 2  & 0  & 0  & 0 \\
DeepSeek-V4-Flash  & 0  & 0  & 0  & 0 \\
GLM-5.2            & 0  & 0  & 0  & 0 \\
\midrule
\multicolumn{5}{c}{Ring} \\
\midrule
Qwen3.5-4B         & 2  & 78 & 10 & 1 \\
Qwen3.5-9B         & 11 & 0  & 5  & 1 \\
Qwen3.6-27B        & 1  & 0  & 0  & 0 \\
Qwen3.6-35B-A3B    & 5  & 4  & 4  & 1 \\
Qwen3.8-27B        & 2  & 0  & 2  & 0 \\
Gemma4-31B         & 0  & 0  & 0  & 0 \\
Devstral-2-24B     & 0  & 0  & 0  & 0 \\
gpt-oss-20b        & 1  & 0  & 0  & 0 \\
gpt-oss-120b       & 1  & 1  & 0  & 0 \\
DeepSeek-V4-Flash  & 0  & 3  & 0  & 1 \\
GLM-5.2            & 0  & 0  & 0  & 1 \\
\bottomrule
\end{tabular}

\caption{Additional agent-level error statistics across different topologies and models. Left: Chain and Star; right: Bipartite and Ring. \textit{Invalid Send} denotes attempts to send to non-neighbors or self. \textit{Corrupted Cipher} indicates ciphertext distortion during forwarding. \textit{Unauth. Decrypt} measures unauthorized decryption attempts (decrypting a cipher without the correct decoder). \textit{Unsent Decrypt} marks successful decryptions that were not forwarded to the owner.}
\label{tab:additional-errors}
\end{table*}

\begin{table*}[ht!]
\centering
\setlength{\tabcolsep}{8pt} 
\resizebox{0.9\textwidth}{!}{ 
\begin{tabular}{llcccc}

\toprule
\textbf{Topology} & \textbf{Model} & \textbf{\shortstack{Misrouted Msgs}} & \textbf{\shortstack{Dropped Request}} & \textbf{\shortstack{Duplicate Msg}} & \textbf{\shortstack{Total Msgs}} \\
\midrule

\multirow{11}{*}{Star} 
 & Qwen3.5-4B      & 48 (27.53\%) & 8 (4.59\%) & 51 (29.06\%) & 174 \\
 & Qwen3.5-9B      & 3 (8.62\%) & 17 (43.97\%) & 0 (0.00\%) & 39 \\
 & Qwen3.6-27B      & 0 (0.00\%) & 0 (0.00\%) & 0 (0.00\%) & 96 \\
 & Qwen3.6-35B-A3B      & 16 (20.85\%) & 8 (9.79\%) & 6 (8.09\%) & 78 \\
 & Qwen3.8-27B      & 0 (0.00\%) & 0 (0.00\%) & 0 (0.00\%) & 96 \\
 & Gemma4-31B      & 0 (0.00\%) & 0 (0.00\%) & 0 (0.00\%) & 96 \\
 & Devstral-2-24B      & 3 (2.87\%) & 3 (2.87\%) & 6 (5.73\%) & 105 \\
 & gpt-oss-20b      & 0 (0.00\%) & 24 (92.21\%) & 0 (0.00\%) & 26 \\
 & gpt-oss-120b      & 0 (1.12\%) & 22 (75.28\%) & 1 (3.37\%) & 30 \\
 & DeepSeek-V4-Flash      & 0 (0.00\%) & 0 (0.00\%) & 0 (0.00\%) & 96 \\
 & GLM-5.2      & 0 (0.00\%) & 0 (0.00\%) & 0 (0.00\%) & 96 \\
\midrule

\multirow{11}{*}{Bipartite} 
 & Qwen3.5-4B      & 45 (26.11\%) & 6 (3.68\%) & 43 (24.95\%) & 172 \\
 & Qwen3.5-9B      & 20 (20.70\%) & 4 (4.56\%) & 7 (7.72\%) & 95 \\
 & Qwen3.6-27B      & 0 (0.00\%) & 0 (0.00\%) & 0 (0.00\%) & 99 \\
 & Qwen3.6-35B-A3B      & 13 (12.12\%) & 0 (0.00\%) & 1 (1.21\%) & 110 \\
 & Qwen3.8-27B      & 26 (21.67\%) & 0 (0.00\%) & 0 (0.00\%) & 120 \\
 & Gemma4-31B      & 0 (0.00\%) & 0 (0.00\%) & 0 (0.00\%) & 100 \\
 & Devstral-2-24B      & 4 (1.71\%) & 2 (1.09\%) & 3 (1.40\%) & 214 \\
 & gpt-oss-20b      & 0 (0.00\%) & 24 (85.71\%) & 0 (0.00\%) & 28 \\
 & gpt-oss-120b      & 0 (0.00\%) & 22 (66.00\%) & 0 (0.00\%) & 33 \\
 & DeepSeek-V4-Flash      & 0 (0.00\%) & 0 (0.00\%) & 0 (0.00\%) & 100 \\
 & GLM-5.2      & 0 (0.00\%) & 0 (0.00\%) & 0 (0.00\%) & 100 \\
\bottomrule
\end{tabular}
}

\caption{Message-level communication errors on the Star and Bipartite topologies; complements Table~\ref{tab:message-level-errors}, which reports Chain and Ring. Misrouted Msgs combines ciphertexts sent away from the decoder holder and decryptions sent away from the ciphertext owner. Dropped Request indicates unactioned \texttt{forward} or \texttt{decrypt\_request} calls. Duplicate Msg indicates looped or repeated messages. Total Msgs reports the average total messages sent per run.}\label{tab:msg-errors-sb}
\end{table*}

\begin{table*}[ht!]
\centering
\small
\setlength{\tabcolsep}{1.75pt}
\begin{tabular}{llcccccc}

\toprule
\textbf{Topology} & \textbf{Model} & \shortstack{Completed} & \shortstack{Init Fail} & \shortstack{Transit to Dec} & \shortstack{Transit to Own} & \shortstack{Decoder Stuck} & \shortstack{Verify Stuck} \\
\midrule

\multirow{11}{*}{Star} 
 & Qwen3.5-4B      & 0.12$_{\pm 0.00}$ & 0.15$_{\pm 0.05}$ & 0.43$_{\pm 0.18}$ & 0.13$_{\pm 0.05}$ & 0.07$_{\pm 0.02}$ & 0.11$_{\pm 0.10}$ \\
 & Qwen3.5-9B      & 0.01$_{\pm 0.02}$ & 0.00$_{\pm 0.00}$ & 0.65$_{\pm 0.46}$ & 0.07$_{\pm 0.12}$ & 0.27$_{\pm 0.32}$ & 0.00$_{\pm 0.00}$ \\
 & Qwen3.6-27B      & 1.00$_{\pm 0.00}$ & 0.00$_{\pm 0.00}$ & 0.00$_{\pm 0.00}$ & 0.00$_{\pm 0.00}$ & 0.00$_{\pm 0.00}$ & 0.00$_{\pm 0.00}$ \\
 & Qwen3.6-35B-A3B      & 0.27$_{\pm 0.29}$ & 0.00$_{\pm 0.00}$ & 0.44$_{\pm 0.46}$ & 0.29$_{\pm 0.51}$ & 0.00$_{\pm 0.00}$ & 0.00$_{\pm 0.00}$ \\
 & Qwen3.8-27B      & 1.00$_{\pm 0.00}$ & 0.00$_{\pm 0.00}$ & 0.00$_{\pm 0.00}$ & 0.00$_{\pm 0.00}$ & 0.00$_{\pm 0.00}$ & 0.00$_{\pm 0.00}$ \\
 & Gemma4-31B      & 1.00$_{\pm 0.00}$ & 0.00$_{\pm 0.00}$ & 0.00$_{\pm 0.00}$ & 0.00$_{\pm 0.00}$ & 0.00$_{\pm 0.00}$ & 0.00$_{\pm 0.00}$ \\
 & Devstral-2-24B      & 0.39$_{\pm 0.31}$ & 0.43$_{\pm 0.22}$ & 0.00$_{\pm 0.00}$ & 0.04$_{\pm 0.07}$ & 0.15$_{\pm 0.12}$ & 0.00$_{\pm 0.00}$ \\
 & gpt-oss-20b      & 0.04$_{\pm 0.00}$ & 0.01$_{\pm 0.02}$ & 0.91$_{\pm 0.02}$ & 0.00$_{\pm 0.00}$ & 0.04$_{\pm 0.00}$ & 0.00$_{\pm 0.00}$ \\
 & gpt-oss-120b      & 0.08$_{\pm 0.04}$ & 0.00$_{\pm 0.00}$ & 0.89$_{\pm 0.02}$ & 0.00$_{\pm 0.00}$ & 0.03$_{\pm 0.02}$ & 0.00$_{\pm 0.00}$ \\
 & DeepSeek-V4-Flash      & 1.00$_{\pm 0.00}$ & 0.00$_{\pm 0.00}$ & 0.00$_{\pm 0.00}$ & 0.00$_{\pm 0.00}$ & 0.00$_{\pm 0.00}$ & 0.00$_{\pm 0.00}$ \\
 & GLM-5.2      & 1.00$_{\pm 0.00}$ & 0.00$_{\pm 0.00}$ & 0.00$_{\pm 0.00}$ & 0.00$_{\pm 0.00}$ & 0.00$_{\pm 0.00}$ & 0.00$_{\pm 0.00}$ \\
\midrule

\multirow{11}{*}{Bipartite} 
 & Qwen3.5-4B      & 0.05$_{\pm 0.06}$ & 0.32$_{\pm 0.04}$ & 0.27$_{\pm 0.13}$ & 0.11$_{\pm 0.02}$ & 0.20$_{\pm 0.08}$ & 0.05$_{\pm 0.05}$ \\
 & Qwen3.5-9B      & 0.04$_{\pm 0.04}$ & 0.08$_{\pm 0.07}$ & 0.28$_{\pm 0.36}$ & 0.16$_{\pm 0.16}$ & 0.40$_{\pm 0.24}$ & 0.04$_{\pm 0.04}$ \\
 & Qwen3.6-27B      & 0.99$_{\pm 0.02}$ & 0.01$_{\pm 0.02}$ & 0.00$_{\pm 0.00}$ & 0.00$_{\pm 0.00}$ & 0.00$_{\pm 0.00}$ & 0.00$_{\pm 0.00}$ \\
 & Qwen3.6-35B-A3B      & 0.64$_{\pm 0.32}$ & 0.01$_{\pm 0.02}$ & 0.04$_{\pm 0.07}$ & 0.31$_{\pm 0.27}$ & 0.00$_{\pm 0.00}$ & 0.00$_{\pm 0.00}$ \\
 & Qwen3.8-27B      & 1.00$_{\pm 0.00}$ & 0.00$_{\pm 0.00}$ & 0.00$_{\pm 0.00}$ & 0.00$_{\pm 0.00}$ & 0.00$_{\pm 0.00}$ & 0.00$_{\pm 0.00}$ \\
 & Gemma4-31B      & 1.00$_{\pm 0.00}$ & 0.00$_{\pm 0.00}$ & 0.00$_{\pm 0.00}$ & 0.00$_{\pm 0.00}$ & 0.00$_{\pm 0.00}$ & 0.00$_{\pm 0.00}$ \\
 & Devstral-2-24B      & 0.27$_{\pm 0.18}$ & 0.60$_{\pm 0.28}$ & 0.00$_{\pm 0.00}$ & 0.00$_{\pm 0.00}$ & 0.13$_{\pm 0.17}$ & 0.00$_{\pm 0.00}$ \\
 & gpt-oss-20b      & 0.01$_{\pm 0.02}$ & 0.00$_{\pm 0.00}$ & 0.96$_{\pm 0.04}$ & 0.03$_{\pm 0.05}$ & 0.00$_{\pm 0.00}$ & 0.00$_{\pm 0.00}$ \\
 & gpt-oss-120b      & 0.09$_{\pm 0.05}$ & 0.00$_{\pm 0.00}$ & 0.88$_{\pm 0.04}$ & 0.03$_{\pm 0.02}$ & 0.00$_{\pm 0.00}$ & 0.00$_{\pm 0.00}$ \\
 & DeepSeek-V4-Flash      & 1.00$_{\pm 0.00}$ & 0.00$_{\pm 0.00}$ & 0.00$_{\pm 0.00}$ & 0.00$_{\pm 0.00}$ & 0.00$_{\pm 0.00}$ & 0.00$_{\pm 0.00}$ \\
 & GLM-5.2      & 1.00$_{\pm 0.00}$ & 0.00$_{\pm 0.00}$ & 0.00$_{\pm 0.00}$ & 0.00$_{\pm 0.00}$ & 0.00$_{\pm 0.00}$ & 0.00$_{\pm 0.00}$ \\
\bottomrule

\end{tabular}
\caption{Agent-level task status distribution on the Star and Bipartite topologies; complements Table~\ref{tab:agent-level-errors}, which reports Chain and Ring. Percentages represent the proportion of agents that terminated in each state; state definitions are identical to Table~\ref{tab:agent-level-errors}.}
\label{tab:agent-states-sb}
\vspace{-0.5cm}
\end{table*}

\section{Computational Budget and Infrastructure}
\label{appx:compute}

This section reports the parameter counts of all evaluated models, the total computational budget, and the hardware used for inference.

\paragraph{Model parameters.}
Table~\ref{tab:compute} summarizes the total and active parameter counts for each model. For dense architectures, total and active parameters coincide. For Mixture-of-Experts architectures, only a subset of parameters is activated per forward pass.

\begin{table}[h]
\centering
\small
\begin{tabular}{lrr}
\toprule
Model & Total (B) & Active (B) \\
\midrule
Qwen3.5-4B & 4 & 4 \\
Qwen3.5-9B & 9 & 9 \\
Qwen3.6-27B & 27 & 27 \\
Qwen3.6-35B-A3B & 35 & 3 \\
Qwen3.8-27B & 27 & 27 \\
Gemma4-31B & 31 & 31 \\
Devstral-2-24B & 24 & 24 \\
gpt-oss-20b & 20 & 3.6 \\
gpt-oss-120b & 120 & 5.1 \\
DeepSeek-V4-Flash & 284 & 13 \\
GLM-5.2 & 753 & 40 \\
\bottomrule
\end{tabular}
\caption{Parameter counts for all evaluated models.}
\label{tab:compute}
\end{table}

\paragraph{Computational budget.}
Each configuration is evaluated over 3 independent simulation runs. Because models of different sizes were served on different numbers of GPUs, a single unified GPU-hour figure would not be meaningful; we therefore report wall-clock time per model and an upper bound for the entire study. As a reference point, the full scaling sweep of Gemma4-31B in Free Text mode (3 runs, 4 topologies, $N$ from 3 to 25) required approximately 30 GPU-hours. No model required substantially more time, because wall-clock cost does not scale with parameter count alone: the number of LLM invocations varies widely across runs, and small models such as Qwen3.5-4B issue many more \texttt{send\_message} calls, often reaching the 2{,}048-call cap. Their evaluation therefore takes time comparable to the 27--31\,B dense models.
 In addition to wall-clock time, Table~\ref{tab:appx:cost} reports the average cost of a single run for every model and topology at $N{=}25$, estimated at OpenRouter\footnote{\href{https://openrouter.ai}{openrouter.ai}} list prices, and Table~\ref{tab:appx:tokens} reports the corresponding average per-run input and output token counts.
All experiments for the eleven models completed within one week on a shared pool of at most 10 GPUs, an upper bound of roughly 1{,}700 GPU-hours.

\paragraph{Hardware and inference framework.}
All models are served using \textbf{vLLM}~\citep{kwon2023vllm} on \textbf{$10\times$ NVIDIA H100 80GB} GPUs.
Most models run in \textbf{BF16} precision without any quantization. 
The only exception is \textbf{GLM-5.2}, which is deployed using the \textbf{W4AFP8} mixed-precision quantization format (4-bit weights for MoE experts and 8-bit activations).
Each inference call uses a maximum generation length of 8{,}192 tokens, temperature 1.0, and top-$p$ 0.95.

\section{Inference Cost and Token Statistics}
\label{appx:appendix-cost}

This appendix quantifies the monetary cost and token footprint of CypherMAS runs at $N{=}25$. Table~\ref{tab:appx:cost} reports the average cost of a single run, estimated at OpenRouter list prices, for the eight models with OpenRouter listings; \textit{Qwen3.5-4B} and \textit{Devstral-2-24B} have no listed price and are therefore excluded. For locally served models such as \textit{GLM-5.2}, the reported cost is an estimate of the equivalent API cost rather than actual expenditure. Table~\ref{tab:appx:tokens} reports average per-run input and output token counts.

Cost is governed by topology rather than model size. For every model, Chain and Ring runs cost roughly an order of magnitude more than Star and Bipartite runs (up to 55$\times$ for \textit{Qwen3.5-9B}), tracking the transmission minima (624/600 vs.\ 96/100): the 753B \textit{GLM-5.2} on Star (\$0.39) is cheaper than the 9B \textit{Qwen3.5-9B} on Chain (\$1.16). Input tokens exceed output tokens by one to two orders of magnitude (e.g., 22.3M vs.\ 238K for \textit{GLM-5.2} on Chain in Strict mode), because every execution hop re-processes the agent's full conversation history. Runs that loop toward the 2{,}048-transmission cap therefore dominate the totals: \textit{Qwen3.6-35B-A3B} accumulates 117.6M input tokens on Ring in Free Text and \textit{Qwen3.5-4B} 77.0M on Ring in Strict mode, an order of magnitude more than capable models on the same topologies.

Transmissions map to cost only loosely. \textit{DeepSeek-V4-Flash} on Ring issues \emph{more} transmissions in Free Text than in Strict mode (691 vs.\ 600) yet costs \emph{less} (\$0.27 vs.\ \$0.45), because its Strict-mode runs carry more input tokens per transmission (10.5K vs.\ 6.2K); \textit{GLM-5.2} on Chain shows the converse, with transmissions and cost falling together (1{,}248$\to$573, \$12.80$\to$\$2.45).

Raw cost must be read jointly with SR, since failed runs terminate early and are correspondingly cheap: the \textit{gpt-oss} models are the least expensive in every configuration precisely because they deadlock. Conversely, the two most expensive configurations in the study --- \textit{GLM-5.2} on Chain in Strict mode (\$12.80) and \textit{Qwen3.6-27B} on Chain and Ring in Free Text (\$12.67, \$12.59) --- are exactly the redundancy pathologies: transmission redundancy translates directly into monetary cost. Normalized by completed agents, \textit{Qwen3.6-27B} on Chain in Free Text is the least cost-effective configuration (\$1.37 per completed agent), while \textit{DeepSeek-V4-Flash} completes agents for one to three cents in both modes.
\begin{table*}[t!]
  \centering
  \setlength{\tabcolsep}{4pt}
  \begin{tabular}{@{} l cccc @{}}
    \toprule
    \textbf{LLM} & \textbf{Chain} & \textbf{Star} & \textbf{Bipartite} & \textbf{Ring} \\
    \midrule
    \multicolumn{5}{c}{\textbf{Strict (Tool-Based) Communication}} \\ \hline
    Qwen3.5-9B       & 1.164 & 0.021 & 0.055 & 0.381 \\
    Qwen3.6-27B      & 2.897 & 0.294 & 0.335 & 3.393 \\
    Qwen3.6-35B-A3B  & 0.408 & 0.061 & 0.096 & 0.581 \\
    Gemma4-31B       & 0.546 & 0.071 & 0.085 & 0.660 \\
    gpt-oss-20b      & 0.067 & 0.008 & 0.012 & 0.067 \\
    gpt-oss-120b     & 0.148 & 0.015 & 0.021 & 0.132 \\
    DeepSeek-V4-Flash  & 0.318 & 0.054 & 0.064 & 0.452 \\
    GLM-5.2          & 12.804 & 0.386 & 0.422 & 3.193 \\
    \midrule
    \multicolumn{5}{c}{\textbf{Free Form Communication}} \\ \hline
    Qwen3.5-9B       & 3.103 & 0.119 & 0.038 & 1.859 \\
    Qwen3.6-27B      & 12.669 & 0.268 & 0.349 & 12.590 \\
    Qwen3.6-35B-A3B  & 1.815 & 0.033 & 0.117 & 7.582 \\
    Gemma4-31B       & 0.735 & 0.032 & 0.038 & 1.038 \\
    gpt-oss-20b      & 0.044 & 0.005 & 0.006 & 0.044 \\
    gpt-oss-120b     & 0.178 & 0.011 & 0.017 & 0.131 \\
    DeepSeek-V4-Flash  & 0.607 & 0.049 & 0.061 & 0.272 \\
    GLM-5.2          & 2.447 & 0.163 & 0.332 & 4.248 \\
    \bottomrule
  \end{tabular}
  \caption{Average cost of a single run (USD) across graph topologies
  at $N{=}25$, estimated at OpenRouter list prices, averaged over 3
  independent MAS runs per configuration. Unsuccessful runs can be
  cheaper, as stalled or deadlocked runs terminate early and issue
  fewer transmissions.}
  \label{tab:appx:cost}
\end{table*}
\begin{table*}[t!]
  \centering
  \setlength{\tabcolsep}{4pt}
  \begin{tabular}{@{} l *{4}{cc} @{}}
    \toprule
    \multirow{2}{*}{\textbf{LLM}}
    & \multicolumn{2}{c}{\textbf{Chain}}
    & \multicolumn{2}{c}{\textbf{Star}}
    & \multicolumn{2}{c}{\textbf{Bipartite Graph}}
    & \multicolumn{2}{c}{\textbf{Ring}} \\
    \cmidrule(lr){2-3} \cmidrule(lr){4-5} \cmidrule(lr){6-7} \cmidrule(lr){8-9}
    & \textit{In} & \textit{Out}
    & \textit{In} & \textit{Out}
    & \textit{In} & \textit{Out}
    & \textit{In} & \textit{Out} \\
    \midrule
    \multicolumn{9}{c}{\textbf{Strict (Tool-Based) Communication}} \\ \hline
    Qwen3.5-4B      & 48.5M & 529.6K & 2.86M & 86.8K & 1.53M & 87.3K & 77.0M & 674.6K \\
    Qwen3.5-9B      & 14.0M & 325.4K & 0.23M & 20.2K & 0.62M & 42.4K & 4.40M & 220.5K \\
    Qwen3.6-27B     & 6.99M & 400.5K & 0.69M & 44.3K & 0.76M & 54.0K & 8.11M & 480.4K \\
    Qwen3.6-35B-A3B & 4.21M & 281.9K & 0.52M & 49.5K & 0.86M & 76.2K & 6.15M & 390.5K \\
    Gemma4-31B      & 5.37M & 184.6K & 0.68M & 28.8K & 0.79M & 39.5K & 6.60M & 196.1K \\
    Devstral-Small-2-24B & 0.78M & 21.8K & 0.50M & 16.2K & 0.75M & 28.3K & 0.95M & 25.2K \\
    gpt-oss-20b     & 1.56M & 356.9K & 0.16M & 43.4K & 0.20M & 77.3K & 1.53M & 363.7K \\
    gpt-oss-120b    & 2.66M & 402.5K & 0.23M & 46.1K & 0.31M & 68.5K & 1.95M & 431.0K \\
    DeepSeek-V4-Flash & 4.95M & 190K & 678K & 33.9K & 757K & 40.8K & 6.28M & 278K \\
    GLM-5.2         & 22.3M & 237.8K & 0.61M & 26.4K & 0.67M & 29.8K & 5.28M & 152.7K \\
    \midrule
    \multicolumn{9}{c}{\textbf{Free Form Communication}} \\ \hline
    Qwen3.5-4B      & 57.4M & 950.7K & 47.2M & 274.4K & 7.07M & 266.5K & 59.8M & 1189.7K \\
    Qwen3.5-9B      & 37.8M & 601.8K & 1.37M & 65.9K & 0.39M & 52.8K & 22.2M & 653.4K \\
    Qwen3.6-27B     & 34.9M & 1102.2K & 0.29M & 90.8K & 0.48M & 103.0K & 33.9M & 1207.8K \\
    Qwen3.6-35B-A3B & 25.2M & 791.0K & 0.14M & 36.8K & 0.85M & 106.6K & 117.6M & 2433.7K \\
    Gemma4-31B      & 7.15M & 267.8K & 0.24M & 31.5K & 0.28M & 37.9K & 10.0M & 392.3K \\
    Devstral-Small-2-24B & 10.4M & 81.8K & 0.14M & 9.8K & 0.18M & 12.5K & 5.66M & 84.1K \\
    gpt-oss-20b     & 0.62M & 317.5K & 0.05M & 37.6K & 0.05M & 49.7K & 0.66M & 304.2K \\
    gpt-oss-120b    & 3.10M & 499.9K & 0.10M & 43.9K & 0.16M & 72.3K & 1.86M & 444.5K \\
    DeepSeek-V4-Flash & 18.1M & 259K & 269K & 35.0K & 341K & 43.7K & 4.28M & 162K \\
    GLM-5.2         & 3.91M & 160.5K & 0.22M & 24.5K & 0.42M & 57.3K & 6.87M & 252.2K \\
    \bottomrule
  \end{tabular}
  \caption{Average input and output tokens processed per run across
  graph topologies at $N{=}25$, averaged over 3 independent MAS runs
  per configuration. Input tokens accumulate because every agent's
  context includes the full observable graph and conversation history;
  runs that loop until the 2{,}048-transmission cap therefore dominate
  the input-token totals.}
  \label{tab:appx:tokens}
\end{table*}

\end{document}